# ICARUS at FNAL

## Proposal

### The ICARUS Collaboration

M. Antonello[1], B. Baibussinov[2], V. Bellini[4,5], H. Bilokon[6], F. Boffelli[7], M. Bonesini[9], E. Calligarich[8], S. Centro[2,3], K. Cieslik[10], D. B. Cline[11], A. G. Cocco[12], A. Curioni[9], A. Dermenev[13], R. Dolfini[7,8], A. Falcone[7,8], C. Farnese[2], A. Fava[3], A. Ferrari[14], D. Gibin[2,3], S. Gninenko[13], F. Guber[13], A. Guglielmi[2], M. Haranczyk[10], J. Holeczek[15], A. Ivashkin[13], M. Kirsanov[13], J. Kisiel[15], I. Kochanek[15], A. Kurepin[13], J. Łagoda[16], F. Mammoliti[4], S. Mania[15], G. Mannocchi[6], V. Matveev[13], A. Menegolli[7,8], G. Meng[2], G. B. Mills[17], C. Montanari[8], F. Noto[4], S. Otwinowski[11], T. J. Palczewski[16], P. Picchi[6], F. Pietropaolo[2], P. Płoński[18], R. Potenza[4,5], A. Rappoldi[8], G. L. Raselli[8], M. Rossella[8], C. Rubbia[19,14,a], P. Sala[20], A. Scaramelli[20], E. Segreto[1], D. Stefan[1], J. Stepaniak[16], R. Sulej[16], C. M. Sutera[4], D. Tlisov[13], M. Torti[7,8], R. G. Van de Water[17], F. Varanini[3], S. Ventura[2], C. Vignoli[1], H. G. Wang[11], X. Yang[11], A. Zani[7,8], K. Zaremba[18]

1. INFN, LNGS, Assergi (AQ), Italy
2. INFN, Sezione di Padova, 35131 Padova, Italy
3. Dipartimento di Fisica, Università di Padova, 35131 Padova, Italy
4. INFN, Sezione di Catania, Catania, Italy
5. Dipartimento di Fisica, Università di Catania, Catania, Italy
6. INFN, Laboratori Nazionali di Frascati (LNF), 00044 Frascati (Roma), Italy
7. Dipartimento di Fisica, Università di Pavia, 27100 Pavia, Italy
8. INFN, Sezione di Pavia, 27100 Pavia, Italy
9. INFN, Sezione di Milano Bicocca, Dipartimento di Fisica G. Occhialini, 20126 Milano, Italy
10. The H. Niewodniczanski Institute of Nuclear Physics, Polish Academy of Science, Kraków, Poland
11. Department of Physics and Astronomy, University of California, Los Angeles, USA
12. INFN, Sezione di Napoli, Dipartimento di Scienze Fisiche, Università Federico II, 80126 Napoli, Italy
13. INR-RAS, Moscow, Russia
14. CERN, Geneva, Switzerland
15. Institute of Physics, University of Silesia, Katowice, Poland
16. National Center for Nuclear Research, Warszawa, Poland
17. Los Alamos National Laboratory, New Mexico, USA
18. Institute for Radioelectronics, Warsaw University of Technology, Warsaw, Poland
19. GSSI, L'Aquila (AQ), Italy
20. INFN, Sezione di Milano, 20133 Milano, Italy

(a) Spokesperson

## Abstract

The Italian Istituto Nazionale di Fisica Nucleare (INFN) and the ICARUS program have originally developed the technology of the LAr-TPC. Exposed in the underground Hall B of the Gran Sasso National Laboratory in Assergi, at 730 km to the neutrino beam from CERN, this large-scale neutrino experiment has been performed with remarkable detection efficiency and it has now successfully completed a three years physics continuous program. It has been a complete success, featuring a smooth operation, high live time, and high reliability. A total of about 3000 CNGS neutrino events have been collected and are being actively analyzed.

In the next two years ICARUS will be considerably improved at CERN with an extensive R&D program. The external dewars of the present T600 detector (760 tons of ultra high purity LAr) will be extensively overhauled and complemented with a similar ¼ scale T150 detector. The R&D improvements are performed in a close collaboration with the LBNE experiment to which the six above INFN Institutions are now participating members. As a main new novelty, a SC magnetic field of about 1 Tesla will be introduced inside the LAr volumes, in analogy to the performance of the traditional bubble chambers.

During 2016 it is proposed to move the whole experiment to FNAL where short base line neutrino beams are available, nicely complementing the already approved MicroBooNe LAr-TPC experiment which will start operation in 2014. The presence of ICARUS at FNAL is an important addition to MicroBooNE since, in the absence of "anomalies", the signals of several detectors at different distances from the beam should be a precise copy of each other for all experimental signatures.

Because of its reduced mass, the anti-neutrino signal in MicroBooNE is too weak for a sensitive comparison. A definitive clarification of the LSND anomaly therefore requires also the exploration of the anti-neutrino signal provided by a much larger T600 mass. The presence of a magnetic field is required in order to separate the anti-neutrino signal in the simultaneous presence of the neutrino, induced background with the help of a dual magnetized baseline arrangement and the T150 at a much shorter distance.

We propose to locate the T600 detector along the Booster Neutrino Beam line (BNB) at an approximate distance of about 700 m; the T150 detector will be located at about 150 ± 50 m from the target. The T600 will also receive in addition a large number ($>10^4$ /year) of $\nu_e$ events from the off-axis kaon-neutrino NUMI beam peaked around about 1 GeV in order to adequately prepare for the LBNE long baseline experiment.

The ICARUS improved program will be operated as an additional element of the wide short baseline neutrino FNAL physics program. Intended primarily in the framework of the preparatory work for the LBNE collaboration, the ICARUS team is also interested in extending the participation to other short baseline neutrino activities collaborating with the already existing FNAL groups.

# Table of contents.

# 1   Introduction.

## *1.1   The development of the LAr-TPC.*

The LAr-TPC technology has been steadily developed over the last thirty years by the ICARUS Collaboration. The peer reviewed scientific publications of the collaboration have today reached a total of 108 papers.

The pre-history of the ICARUS LAr-TPC program can be traced at Harvard University in 1985/88 with the publication of few seminal papers: *E. Aprile, K.L. Giboni and C. Rubbia, "A study of ionization electrons drifting large distances in liquid and solid argon"* (*Nucl. Instr. and Meth. A241, (1985), 62*) [1], *J.N. Bahcall, M. Baldo-Ceolin, D.B. Cline and C. Rubbia, "Prediction for a Liquid Argon Solar Neutrino Detector"* (*Phys. Lett. 178B, (1986), 324*) [2].

The first laboratory scale detector developed at CERN and INFN was reported in 1988, *E. Buckley et al., "A study of ionization electrons drifting large distances in liquid argon", (Nucl. Instr. and Meth. A275, (1989), 364)* [3] with Liquid Argon corresponding to an impurity concentration of 30 parts per trillion of Oxygen equivalent.

At about the same time the ICARUS collaboration was launched, having in mind the LNGS Laboratory as the next goal. Proposals were presented and approved by the INFN: The ICARUS Collaboration, ICARUS-I. *"A proposal for the Gran Sasso laboratory"*, Proposal, INFN/AE-85/7, Frascati (Italy, 1985); ICARUS-II. *"A second generation proton decay experiment and neutrino observatory at the Gran Sasso laboratory"*, Proposal, VOL I (1993) & II (1994), LNGS-94/99; *"A first T600 ton ICARUS detector installed at the Gran Sasso laboratory"*, Addendum to Proposal, LNGS 95-10 (1995) [4].

During the nineties, progressively larger detectors were constructed with the support of INFN and a transition from laboratory to industry was carried out. The ICARUS development program culminated in 2001 with the successful operation on surface in Pavia with Cosmic Rays of the first T300 module, a physics grade detector of a mass of 360 tons and a fiducial volume of about $19 \times 3 \times 3$ m$^3$ and a maximum drift path of 1.5 m [5]. Signals were provided by a set of three parallel planes of wires, 3 mm apart, 3 mm pitch, facing the drift volume, oriented at different angles (0°, +60°, -60°) with respect to the horizontal direction. A second T300 module was constructed in the following two years, bringing the detector to the final T600 configuration of today.

The first detection of real neutrinos from a LAr-TPC was performed at the end of the nineties with a 50 liters ICARUS-like chamber located between the CHORUS and NOMAD experiments at the CERN West Area Neutrino Facility (WANF) and published in 2006, *"Performance of a liquid argon time projection chamber exposed to the CERN West Area Neutrino Facility neutrino beam"* F. Arneodo et al., Physical Review D (2006) (Vol. 74, No. 112001) and on arXiv:physics/0609205v1 [physics.ins-det] [6].

Another important milestone was achieved with the development of the detection of high energy muons in order to complete the kinematical determination of neutrino events: *"Measurement of through-going particle momentum by means of multiple scattering with the ICARUS T600 TPC"*, A. Ankowski et al. Eur. Phys. J. C 48 (2006) 667-676 [7].

The engineering extension of the experiment to 3000 ton in the Gran Sasso underground Laboratory (LNGS) of the Istituto Nazionale di Fisica Nucleare (INFN) was developed, however not implemented by INFN and superseded by the design of a future detector of a much larger mass. In 2007 the conceptual report of a much larger mass detector called MODULAR was developed, subdivided in individual units of 5 to 10 kton LAr, with remarkable similarities with the LBNE proposal for Homestake [8]. The design has been published as *"A new, very massive modular Liquid Argon Imaging Chamber to detect low energy off-axis neutrinos from the CNGS beam. (Project MODULAr)"*; B. Baibussinov et al., arXiv:0704.1422v1 [hep-ph] 11 Apr 2007 [9] and B. Baibussinov et al. Astroparticle Physics 29 (2008) 174–187 [10].

The T600 detector was moved in 2008 to the Hall B of LNGS in association with the neutrino beam coming from CERN. The new, demanding problems associated with the underground operation of such a large detector (Figure 1) were successfully solved.

T600 was finally operated from 2010, collecting cosmic ray data and neutrino events from the CERN Neutrino to Gran Sasso (CNGS) beam in the physics experiment CNGS2 [11].

## 1.2  *Early physics with the ICARUS experiment.*

The ICARUS experiment has performed with a remarkable detection efficiency and it has now successfully completed a three years physics program at CNGS. The T600 was exposed to the CNGS beam. From October 2010 to December 2012, neutrino events have been collected, corresponding to $8.6 \times 10^{19}$ protons on target with an efficiency exceeding 93%. Additional data were also collected with cosmic rays to study the detector capability for atmospheric neutrinos and proton decay.

From the technological point of view, the T600 run was a complete success, featuring a smooth operation, high live time, and high reliability. A total of about 3000 CNGS neutrino events have been collected and are being presently actively analyzed.

Some early findings from the LNGS program are hereby briefly described. Several papers have been already published with peer reviews:
- *"Underground operation of the ICARUS T600 LAr-TPC: first results"*, JINST 6 (2011) P07011 [11].
- *"A search for the analogue to Cherenkov radiation by high energy neutrinos at superluminal speeds in ICARUS"*, PLB 711 (2012) 270 [12].
- *"Measurement of neutrino velocity with the ICARUS detector at the CNGS beam"*, PLB 713 (2012) 17 [13].

- *"Precision measurement of the neutrino velocity with the ICARUS detector in the CNGS beam"*, JHEP 11 (2012) 049 [14].
- *"Precise 3D Reconstruction Algorithm for the ICARUS T600 Liquid Argon Time Projection Chamber Detector"*, AHEP 2013 (2013) 260820 [15].
- *"Experimental search for the LSND anomaly with the ICARUS detector in the CNGS neutrino beam"*, EPJ C73 (2013) 2345 [16].
- *"Search for anomalies in $\nu_e$ appearance from $\nu_\mu$ beam"*, Eur. Phys. J. C (2013) 73:2599 [17].

Besides physics results, many verifications of the detector performances, validations of the simulation packages and progresses in event reconstruction have been studied. For instance, the *dE/dx* of muons, protons and pions is well reproduced by simulations, as well as the energy deposit by neutrino interactions [16]. Also, the unambiguous capability to discriminate electrons from photons, due to the remarkable quality of the events in LAr-TPC has been extensively verified with actual neutrino data.

Of particular interest is the search for an anomalous, LSND like [18] appearance of $\nu_e$ on a sample of 1995 events, out of which 4 were identified as $\nu_e$ CC, with an expected background from conventional sources of (6.4 ± 0.9) events [17]. The corresponding new limits on the oscillation probability are $P(\nu_\mu \rightarrow \nu_e) = 3.4 \times 10^{-3}$ and $P(\nu_\mu \rightarrow \nu_e) = 7.6 \times 10^{-3}$ at 90% and 99% confidence levels respectively (Figure 2). This result, in contradiction with the best fits of the low energy anomaly of MiniBooNe [19], has been recently confirmed also by the OPERA experiment [20]. Further analysis of the large amount of physics data is progressing, as well as detailed evaluation of the technical aspects.

### *1.3 Observed technical performance of the LAr-TPC.*

The three-year physics run at LNGS, together with all previous test beam runs, has allowed an assessment of the actual detection capabilities of the LAr-TPC.

The availability of the first physics neutrino data has permitted the optimization of the development of the tools and of the data analysis. A number of physical features that will be also relevant for the present FNAL proposal have been obtained from the experimental data.

A novel 3-D reconstruction approach has been developed [15]. It is based on the Polygonal Line Algorithm fit in 3D space, optimized to all available hits in the 2D wire planes and all the identified 3D reference points (interaction vertices, peculiar points like delta rays). The main advantage is that 2D hit-to-hit associations are not needed and therefore missing track parts in single view are accepted as well as horizontal tracks except for pathological situations, such as long missing segments at the track ends. The new approach now implemented may be used for reconstruction of cascade-like objects, as well as precise reconstruction of the vertex region and in global event reconstruction.

The electromagnetic energy resolution in the sub-GeV energy range is $\sigma(E)/E = 0.03/\sqrt{E(GeV)} + 0.01$, in agreement with $\pi^0$ invariant mass measurements [21]. The estimated energy resolution for a hadronic cascade is $\sigma(E)/E = 0.3/\sqrt{E(GeV)}$. However the LAr-TPC detector allows to identify and measure, track by track, each hadron produced in low energy neutrino interactions through ionization and range, leading generally to a much better energy resolution.

The *dE/dx* distribution for CNGS muon tracks recorded in CC events has been compared with the expectations from detailed beam and detector simulations [16]. Tracks are distributed over the whole detector volume and have been recorded during several months of operation, thus including all possible effects due to spatial or temporal non-uniformity. The agreement is at the level of 2.5% on the average value (Figure 3). Similar comparisons have been performed on single tracks from long tracks of stopping muons, in which the energy can be directly measured. The distribution of *dE/dx* for each track, sampled at each wire, has been fitted with a Landau function convoluted with a Gaussian distribution of the electronic noise with a r.m.s. of about 10%, corresponding to the expected signal/noise ratio ~10/1 of the individual signal hits. The fitted value of the most probable *dE/dx* agrees within 2% with the Monte Carlo expectations. A similar agreement has been obtained for protons and pions.

In absence of a magnetic field, the momentum of muons can be determined through the reconstruction of multiple Coulomb scattering inside the detector [7]. This method has been verified on CNGS stopping muons by comparison with the integrated *dE/dx* loss. A typical resolution $\Delta p/p \sim 15\%$ due to multiple scattering has been obtained for energies of few GeV and track lengths of a few meters.

A specific particle Identification algorithm (PId) has been developed allowing for a reduction of the systematic effects due to trajectory and charge measurement errors. Stopping particle identification is based on the dependency of reconstructed *dE/dx* versus track residual range. The observation of decay products allows perfect separation (close to 100%) between protons and kaons.

A set of stopping particles has been selected, to check the identification algorithm. The PId clearly separates the proton tracks from the pion and muon sample. Corrections for the effect of quenching are then applied to each energy deposition according to the Birks law, bringing the resulting *dE/dx* versus range patterns in agreement with the Bethe-Bloch formula. Kaons have also been selected in the data. An example of kaon decay and corresponding PId patterns is shown in Figure 4.

Due to the excellent imaging capability of the LAr-TPC, $\pi^0$ from NC events are easily distinguished from electrons. Events where both photon conversion points can be distinguished from the interaction vertex can be rejected. Photons are immediately identified if the conversion distance is exceeding 2 cm. The remaining $\pi^0$ background is further reduced by discarding events where the parent $\pi^0$ mass can be reconstructed. Only few % of $\pi^0$ survives these cuts. The remaining photons can be discriminated from electrons on the basis of *dE/dx* losses. Since photons convert mainly through pair production, the ionization at the beginning of the shower corresponds to that of two electrons. The possible photon misidentification is essentially due to photons undergoing Compton scattering, whose cross section becomes negligible with respect to the pair production above a few hundreds MeV. Monte Carlo studies indicate in CNGS events a residual contamination of about 0.18% for the energy spectrum of photons from pion decays, rising to a few percent in the sub-GeV energy region. The loss in efficiency for electron showers is only 10%.

First results from an ongoing study on low energy showers from isolated secondary $\pi^0$'s in the T600 CNGS data confirm the MC expectation (Figure 5). Showers have been reconstructed in 3D with the same algorithm used for tracks. The effectiveness of e-$\pi^0$ separation in the ICARUS LAr-TPC has been demonstrated even in the heavily crowded high-energy CNGS events (Figure 6) allowing to obtain a new limit on $\nu_\mu \rightarrow \nu_e$ oscillations, that has been recently published [17].

### 1.4  The potentialities of ICARUS at FNAL.

The presently completed experiment at LGNS with the neutrino beam from CERN had been based on a sample of about 3000 events. The future experiments with the short baseline arrangement at FNAL will however collect a sample of many millions of events. The automation of the event reconstruction is a mandatory, challenging task, given the very large expected data sample. An algorithm to identify and reconstruct the primary vertices is under test on the CNGS neutrino CC interactions. Vertices identified in two views are merged together to reconstruct a 3D vertex. Preliminary value of the efficiency of this algorithm on real CNGS data is close to 97% (in 3% of the cases the primary vertex is found at more than 100 mm from the visually identified one). Tools for automatic event reconstruction are under evaluation.

The introduction of an adequate magnetic field to our large LAr-TPC will represent another major improvement complementing the calorimetric tracking of the events with the direct measurement of the sign of the charge and of the momentum by curvature of the tracks, in a complete analogy of the performance of the traditional bubble chamber.

Besides the specific addition of the magnetic field to the LAr-TPC several other activities are foreseen, which will be completed at CERN: (1) an overhaul of the T600, (2) the construction of a similar but smaller T150 detector and (3) a number of R&D's improvements on the technology of the LAr-TPC.

Such a preparatory phase will be completed within about two years. According to the present proposal, it is envisaged to move the whole ICARUS experimental setup to FNAL early in 2016 where the detectors will be operated as an additional element of the short baseline neutrino physics program. Intended primarily in the framework of the preparatory work for the LBNE collaboration [8], the ICARUS team is also interested in extending the participation to other short baseline neutrino activities collaborating with the already existing groups.

The T600 detector should be located along the Booster Neutrino Beam line (BNB [22]) at an approximate distance of about 700 m from the target. The detector will not only record events from the BNB but it will also receive neutrinos from the off-axis kaon-neutrino NUMI beam [23], peaked around about 2 GeV and with a flux of $\nu_e$ events as large as about 5%.

The presence of the ICARUS detectors T600 and of the new T150 [24] to be built will complement the information coming from the already approved MicroBooNE LAr-TPC detector [25] now foreseen to start operation in 2014 at the distance of 470 m from the target.

As well known, the previous MiniBooNE experiment has not entirely confirmed the presence of the LNSD signal above 480 MeV neutrino energies, but it has introduced a new low energy effect at smaller energies [19]. The ICARUS and OPERA experiments [17],[20] but at a much larger L/E have already concluded on the instrumental nature of this new MiniBooNE low energy anomaly. Their actual origin will be further studied by the MicroBooNE detector, which will unambiguously identify electrons from single photons.

However, as described in more detail later (see for instance Figure 27), if operated as a single LAr-TPC detector, MicroBooNE is seriously limited by the presence of instrumental effects. The $\nu_e$ intrinsic background error, in the configuration with a single detector experiment, is ~15% mainly due to systematic uncertainties on kaon production at the relevant energies, estimated to be at the 30% level [26].

A proposal for a dual baseline experiment in which events are detected simultaneously at different distances from the target has been initially presented at CERN as early as 2009. The simultaneous presence of ICARUS at FNAL is an important addition to MicroBooNE since, in absence of "anomalies", the signals of several detectors at different locations should be a precise copy of each other for all experimental signatures of neutrino oscillations without any need of Monte Carlo comparisons.

Because of its mass, MicroBooNE is necessarily limited to the study of neutrinos, the antineutrino signal being too weak for a sensitive comparison. On the other hand, the LSND signal is primarily related to anti-neutrinos. A definitive clarification of the LSND anomaly therefore requires also the exploration of the signal from anti-neutrino that ICARUS may provide. The weakness of the anti-neutrino signal is compensated by the much larger mass of the T600. However the presence of the magnetic field is also necessary in order to separate the anti-neutrino events from a possible LSND antineutrino-anomaly free from the much larger background due to the simultaneous presence of the neutrino induced background of the BNB beam. In order to determine unambiguously the presence of the LSND oscillations in the antineutrino signal with a dual baseline arrangement, a second near detector could be perfected with the help of the (magnetized) T150 at a much sorter distance.

As well known, the search for CP violation for the future LBNE long baseline detector is based on the observation of a small sample of $\nu_e$ events starting from an initial $\nu_\mu$ beam. This requires extensive tests of the expected performances of these events in a LAr-TPC. Operated in the off-axis kaon-neutrino NUMI beam, the ICARUS T600 will permit to collect a very large (> $10^4$/year) number of unbiased $\nu_e$ events that can be used in order to perfect the analysis programs of the subsequent long baseline LBNE detector, measuring with precision the cross sections and the specific identification patterns of these events. Of particular interest are the $\nu_e$ events in the region ≤ 2 GeV, where, due to the presence of CP-violation, it is expected a dip and of a subsequent second peak, in an energy domain in which competing backgrounds have still to be conclusively mastered.

On a longer timescale, NUSTORM [27] has been proposed in order to deliver beams of neutrino and antineutrino from the decay of a stored muon beam. A potential sterile neutrino signal —both in the appearance and disappearance modes— is detected by comparing the signals of the wrong sign neutrino-mu at two different distances from the production target. We believe that our pair of magnetized LAr-TPCs ICARUS detectors with a field of the order of 1 Tesla may constitute an interesting alternative or a complement to the presently chosen magnetized iron detectors.

## 2   A description of the ICARUS detectors.

### 2.1   *The T600 detector.*

The ICARUS T600 detector [5] is made of a large cryostat split into two identical, adjacent modules with internal dimensions $3.6 \times 3.9 \times 19.6$ m$^3$ filled with about 760 tons of ultra-pure liquid Argon (Figure 7). Such units may be operated together as a unique detector.

Each module houses two TPCs separated by a common central cathode. A uniform electric field ($E_D$ = 500 V/cm) is applied to the drift volume. Each TPC is made of three parallel wire planes, 3 mm apart, with 3 mm pitch, facing the drift path (1.5 m). Globally, 53248 wires with length up to 9 m are installed in the detector. By appropriate voltage biasing, the first two signal sensing planes (Induction-1 and Induction-2) provide induced signals in a non-destructive way, whereas the last Collection plane finally collects the ionization charge. The reliable operation of the high-voltage system has been extensively tested up in the ICARUS T600 up to about twice the operating voltage (150 kV, corresponding to $E_D$ = 1 kV/cm).

On each chamber, the wire planes are oriented at 0°, ±60° angles with respect to the horizontal direction. Therefore a three-dimensional image of the ionizing event is reconstructed combining the wire coordinate on each plane at a given drift time. A remarkable resolution of about 1 mm$^3$ is uniformly achieved over the whole active volume (~340 m$^3$ corresponding to 476 t).

The measurement of the absolute time of the ionizing event, combined with the electron drift velocity information ($v_D$ ~1.6 mm/μs at $E_D$ = 500 V/cm), provides the absolute position of the track along the drift coordinate. The absolute time of the ionizing event is given by the prompt detection of the scintillation light produced in LAr by ionizing particles. The detection of VUV scintillation light ($\lambda$ = 128 nm), at LAr cryogenic temperature is ensured by photosensitive layers installed behind each of the wire planes and coated with wavelength shifter.

All the detector components are held by a self-supporting low carbon stainless steel structure. The structure is coupled to the floor of the aluminum vessels by 10 supports. One of them is fixed, while the others are sliding to allow for the relative movement of the detector structure and the aluminum vessels during the cooling phase due to thermal shrinking.

Stainless-steel chimneys are aligned in two rows on the aluminum ceiling of each module (18 per row) and terminated with special vacuum-tight feed-through flanges (INFN patent RM2006A000406). They ensure the passage of the wire signal cables (576 per flange) and other detector subsystems and the control of the instrumentation. Each flange is connected to a single electronic rack where the front-end electronics, the digitizers and the memory buffers are hosted for the readout of 576 channels.

Electronics was designed to allow the continuous read-out, digitization and independent waveform recording of the signals from each wire of the TPC [28], [29]. The read-out architecture consists of a front-end low noise charge sensitive pre-amplifier, allowing a signal-to-noise better than 10:1. Signals coming from each wire are independently digitized every 400 ns with the help of a 10 bit FADC. This scheme is implemented on a single VME-like analogue board, hosting 32 channels amplifiers, multiplexers, ADC's and a 20-bit, 40 MHz serial link that conveys the information to a digital board for filtering and buffering.

A thermal insulation, assembled to realize a tight containment, surrounds the two cold vessels. To maintain the cryostats bulk temperature of 89 K within 1 K, a thermal shield with boiling Nitrogen circulating inside, is inserted between the insulation and the aluminum cold vessels.

Nitrogen, used to cool the whole T600, is stored in two 30 m$^3$ LN$_2$ tanks. Its temperature is fixed by the equilibrium pressure in the tanks (~2.1 bar, corresponding to about 84 K), which is kept stable in a steady state. A dedicated re-liquefaction system is installed ensuring a safe operation in closed-loop. A total of twelve cryo-coolers with 48 kW total cold power are presently available and they will be sufficient for both T600 and T150 detectors.

## 2.2 The T150 detector.

The T150 detector will be similar to one fourth of T600, with the length reduced to 10 m and a total LAr mass of 200 t (119 t active mass) (Figure 8) [30]. A revised design of wire chambers mechanics will be implemented. The construction and assembly of the T150 will take full advantage of the existing know-how, equipment and tooling developed for the T600, with considerable time and cost savings. Some 15,000 channels of electronics will be built. The T150 detector will also represent an ideal playground for the LAr-TPC improvement program (described in Section 3) in view of the LBNE experiment.

## 2.3 LAr purification systems and cryogenic plant.

Argon purification represents one of the most fundamental issues for a successful initial development and subsequent operation of a LAr-TPC. The present elaborated system is the result of many years of experience ensuring that the residual contamination of Oxygen equivalent (electron capture cross-section) can be stably maintained at the level of purity of the order of few tens of ppt (parts per trillion).

The LAr purification solution developed by ICARUS over about 20 years of R&D is essentially based on three key elements: (1) Use of commercial filters, carefully selected among the many available on the market; (2) High Vacuum techniques and protocols; (3) Continuous argon purification through recirculation both in the liquid and in the gas phase, which is essential for the detector operation over long time periods.

It is worth mentioning that in practice the purity is quickly lost already by few days of incomplete operation. The flawless reliability of the whole system is therefore of fundamental importance for the operation of the plant and an adequate performance of the detector. The free electron lifetime in the T600 has been presently maintained at the level of about 6 ms [11], corresponding to the loss of 17% if the signal at the maximum drift distance of 1.5 m (Figure 9).

Extremely high purity values have been already measured with cosmic muons in the ICARINO small prototype [31]. The result reported is $\tau_{ele} \approx 21$ ms corresponding to ~15 ppt, namely a ~$10^{-11}$ molecular Oxygen equivalent impurities. This result demonstrates the suitability of the Oxysorb-Hydrosorb filtering method for very high purification levels. R&D work on laboratory tests has indicated that a lifetime exceeding about 20 ms may be achievable in the future also on a large scale.

Each T600 module (Figure 10) is equipped with: (1) two gaseous (GAr) recirculation units with a total flow rate of 50 Nm$^3$/h followed by a condenser and an Oxysorb filter; the reliquified Ar is re-injected at the top of the LAr volume; (2) one LAr recirculation unit working at a total nominal flow rate of 2 liquid m$^3$/h, followed by 4 Hydro/Oxysorb filters fed in parallel; LAr is re-injected at the opposite side (about 20 m apart) of the vessel at the bottom of the LAr volume.

The GAr recirculation system also performs as a pressure controller. Heat entering from the top through the cables, flanges, etc., is only partially dumped by the thermal shield and it produces a surface evaporation of the LAr.

GAr re-condensation power is provided by LN$_2$ forced flow into a heat exchanger placed at the entrance of the GAr recirculation systems. The LAr recirculation system is cooled by an extension of the same forced LN$_2$ cooling circuit (for the argon pump, the filters and the return line) to prevent gas production that would significantly reduce the mass flow and which will be detrimental for the HV system.

The principal cooling power is provided by circulation into the thermal shield of liquid and gas nitrogen mixture. Forced circulation has been adopted with a centrifugal pump as standard operating mode. The present installation allows also the operation of the cooling system in pure passive mode by means of $LN_2$ gravity circulation through the thermal shields and additional gas condensers for pressure control. This ensures the cryogenic operation of ICARUS, without the argon purifiers, especially in case of long blackouts or $LN_2$ pumps failure.

The $LN_2$ maintains the requested uniform temperature of the LAr. The cooling capacity needed to re-condense the gaseous Nitrogen generated by the entire system is provided by Stirling cryo-coolers (Figure 11), each producing about 4 kW of cold power at 80 K. The expected consumption rate is of about 14 kW for the T600 and of about 6 kW for the T150. Therefore the expected global $LN_2$ consumption may be conservatively handled with the help of a fraction of the 12 Stirling units already presently available at LNGS.

The future extension of the present method to an amount of LAr equal to 70 times the ICARUS-T600 volume as required by the long baseline LBNE detector with 35 kton of LAr demands a number of important modifications that will have to be studied and to which the ICARUS team may contribute significantly.

The purification system will have to be enhanced in view of its application non-evacuated larger size cryostats. The uniformity of lifetime over the volume needs to be ensured with a suitable LAr distribution system. The recirculation flow rate must be vastly increased and all materials to be used must be previously selected and certified according to their contribution to the purity.

## *2.4 Preparatory work at CERN.*

The ICARUS T600 detector is in the process of being provisionally moved to CERN with minimal changes, preserving most of the existing and already operational equipment. In view of its new operation and since the detector has been constructed more than ten years ago, some components are to be renewed: cold vessels, thermal shields and external insulation, photo-detectors instead of photomultipliers in view of the additional presence of the magnetic field.

It is planned to introduce an appropriate magnetic field in the direction orthogonal both to the incoming beam and to the drift direction. The coils will be superconducting either operating at 4.2 K and liquid He with ordinary copper stabilized Nb-Ti or at 20 K with gaseous He and with the use of a new superconducting cable made with $MgB_2$ and recently developed at CERN.

The T600 TPC internal detectors will be extracted fully assembled from the cryostats at LNGS into a clean room container protecting and supporting the chambers for the transport to CERN. Several components of the detector and of the associated cryogenic equipment must be disassembled, like for instance, the cryogenic storage tanks, the auxiliary plants and the $LN_2$ re-liquefaction system. Also the present electronics housed in 96 racks on top, will be dismounted. The whole detector will be re-assembled at CERN before the transport to FNAL. At least at this stage no cryogenic operation of the T600 is foreseen at CERN.

The new T150 detector will be realized as an exact "clone" of one T600 module, with a reduction of a factor 2 (about 10 m) in length. The additional LAr mass will be of 200 t (119 t active mass). The same design of wire chambers, mechanics, wire planes and the 1.5 m maximum drift distance will be preserved. The cold vessel, the insulation vessel and the cryogenics equipment will also be consistent with the solutions already adopted for the overhauled T600.

Some 15000 new channels of electronics shall be built for the T150. The basic architecture will be the same as the one of T600, taking advantage of more modern components. It is likely that in order to test the new detector operation at the cryogenic temperature and filling of LAr is likely required already at CERN.

## 2.5 *The new cold vessels, thermal shields and passive insulation.*

LAr will be contained in three mechanically independent vessels, two of about 270 m$^3$ each for the T600 detector and one of 165 m$^3$ for the T150. According to the present experience, to outgas efficiently the internal surfaces and obtain an appropriate LAr purity, the cold vessels must be evacuated to less than $10^{-3}$ mbar. Therefore the vessels need to stand the vacuum and to be tight to better than $10^{-5}$ mbar l sec$^{-1}$. Moreover they will be designed for a maximum operating internal overpressure of 1 bar (0.45 bar relief valve settings in addition to 0.55 bar hydrostatic pressure).

The new T600 vessels will be of parallelepiped shape with internal dimensions 3.6 (w) × 3.9 (h) × 19.6 (l) m$^3$ matching exactly the existing TPC internal detector. The T150 vessel will have the same cross-section (internal 3.6 (w) × 3.9 (h) m$^2$; external 3.9 (w) × 4.2 (h) m$^2$) and an internal length of 11.8 m (12.1 m external).

They will be realized by welding together extruded aluminum profiles with a significant simplification with regard to the aluminum honeycomb used in the present detector, at the cost of a slight increase in the cryostat weight, 30 t each (Figure 12). The use of aluminum LAr vessels is also particularly attractive in view of the very good shielding offered against external electronic noises and the large thermal conductivity that improves the temperature uniformity inside the LAr.

As in the original T600 vessels, the double-layered walls can be evacuated, leading to efficient leak detection and repair. The cold vessels are enclosed inside a common heat exchanger (thermal shield) in which two phases (gas+liquid) nitrogen is circulated (Figure 13). The mass ratio between the liquid and the gas is kept to less than 5:1 in order to ensure temperature uniformity all along the shield.

This shield is an essential component of the detector since it minimizes the heat load to the bulk of the sensitive LAr volume in order to: (1) suppress any risk of LAr boiling that could be detrimental for the HV system; (2) establish very good temperature uniformity ($\Delta T < 1$ K), through LAr de-stratification as required for uniformity of electron drift velocity and LAr purity; (3) reduce the internal temperature gradients during the cool down making it faster. The critical period during which the argon purifier is switched-off is kept to a minimum in order to preserve an acceptable initial LAr purity.

A purely passive insulation is chosen, based on the "Membrane tanks" technology, adapted to the ICARUS design. This technique has been developed for 50 years and is widely used for large industrial storage vessels and ships for liquefied natural gas [32]. It constitutes also the reference choice for the large multi-kton cryostats of the LBNE project in USA.

An insulation thickness of 60 cm will be used for the bottom and lateral sides; for the top-side a maximum thickness of about 40 cm will be used. With this configuration, the expected thermal losses will be less then 10 W/m$^2$, resulting in a heat loss through the insulation of ~6 kW.

As pointed out later, the additional insertion of a magnetic coil in the cryogenic segment with layers of Mylar reflecting Super-Insulation adds about 40 cm to the lateral sides. The cooling introduced between helium and the nitrogen temperatures is relatively small with respect to the total estimated and nitrogen related cooling power and it has not been subtracted to the design parameters.

All the external heat contributions (cables, pumps, transfer lines, etc.) can be accounted for a value not exceeding 5.4 kW leading to a total heat load of about 11.5 kW. The already existing cryo-coolers allow sufficient power to operate also the T150 detector. In Figure 14, the installation layers of all the elements composing the external detector is shown.

## 2.6 *Magnetizing the LAr-TPC.*

A new robust R&D program, aiming at equipping both the T600 and the T150 LAr-TPC's with a magnetic field, will be pursued following the ideas already fully described in the first ICARUS proposal [4] originally presented and approved by the INFN as early as in 1985. As already pointed out, the realization of a magnetized large LAr-TPC will represent a major improvement allowing both the particle charge identification and the momentum measurement, complementing the multiple scattering technique presently used for muons.

The presence of the magnetic field nicely complements the calorimetric tracking of the events, introducing with the charge and curvature of the tracks a complete analogy to the one of the traditional bubble chamber. A typical event is shown in Figure 15 for a 4 GeV neutrino event simulation, decaying into an electron, a $\pi^0$, a $\pi^+$ and a proton.

The introduction in a LAr-TPC of a magnetic field of appropriate intensity has been later discussed by a number of authors [33]. With respect to the main direction of an incoming neutrino beam the magnetic field should be oriented orthogonally to the electric field direction, in the vertical direction, resulting in a track curvature along the drift time plane.

In a TPC the drifting electron trajectories are modified by the presence of the magnetic field. In a simple approximation [34], gas kinetic theory provides the drift velocity $v$ as a function of the mean collision time $\tau$ and the electric field $E$: $v = eE\tau/m_e$ (Townsend's expression). In analogy with the well known case of a gaseous LAr-TPC, a simple theory, the friction force model, provides an expression for the vector drift velocity $v$ as a function of electric and magnetic field vectors $\vec{E}$ and $\vec{B}$, of the Larmor frequency $\omega = eB/m_e$, and of the mean collision time $\tau$:

$$\vec{v} = \frac{e}{m_e} \frac{\tau}{1+\omega^2\tau^2} \left[ \vec{E} + \frac{\omega\tau}{|B|}\left(\vec{E}\times\vec{B}\right) + \frac{\omega^2\tau^2}{|B|^2}\left(\vec{E}\circ\vec{B}\right)\vec{B} \right]$$

As a good approximation, and for moderate fields, one can assume that the energy of the electrons is not affected by $\vec{B}$, and use for $\tau$ the values deduced from the drift velocity at $B = 0$ (the Townsend expression). For $\vec{E}$ perpendicular to $\vec{B}$, the drift angle relative to the electric vector is $tan(\theta_B) = \omega\tau$ and $v = (E/B)\left(\omega\tau/\sqrt{1+\omega^2\tau^2}\right)$.

For the typical ICARUS value of E = 500 V/cm and B = 1 Tesla in LAr at the density of 1.4 g/cm$^3$, the experimental drift velocity is $v = 1550$ m/s, the cyclotron frequency $\omega = 1.76 \times 10^{11}$ s$^{-1}$, the mean collision time $\tau = 2.87 \times 10^{-13}$ s. The corresponding drift angle is then rather small, $\tau_B = 5.05 \times 10^{-2}$ rad = 2.9°. Some experimental data on LAr have been reported [35], however in conditions which differ from the ones of the LAr-TPC. Additional direct measurements are therefore required.

The new T150 LAr-TPC detector to be constructed first will represent a first comprehensive test of the operability of a physics grade LAr-TPC in an adequate magnetic field. The presence of the magnetic field will no longer permit the simultaneous operation with ordinary phototubes. The otherwise very abundant scintillation light (a 100 MeV m.i.p. will emit in LAr as many as $2 \times 10^6$ UV photons at 128 nm) might be recovered using instead solid-state photo-detectors.

The realization of the magnetic field is briefly described. The most practical design would be a solenoid with vertical axis surrounding the cold vessels of the T600 and the T150. This configuration will ensure an adequately uniform magnetic field in the LAr-TPC active volume oriented orthogonally to both the electric field and the beam direction, with track deflection occurring in the plane where the best space resolution is measured. Alternatively a pair of Helmoltz coils could be located at the edges of the LAr volume.

Various alternatives have been considered for the realization of the superconducting coil. The utilization of a Warm Superconductor, with Bi-2223 or YBCO tapes, although directly possible at the operational temperature of the LAr, has been excluded because of its large cost and the present status of the art of this technology. Therefore an additional cooling environment operating with Helium has to be added, inside the thermal insulation vessel together with at the main cryogenic system.

The standard Niobium-Titanium superconducting cable uses liquid Helium at temperatures between 4.2 K and 1.9 K (-268.8 °C and -271.1 °C). For the present application the value of 4.2 K has been chosen. An alternative choice could be based on new Magnesium Diboride ($MgB_2$) SC cable. $MgB_2$ is considerably less expensive than High Temperature Superconductors and offers the major advantage that it remains functional at up to 25 K (-248 °C). The material has been available since the 1950s, but its SC properties were only discovered in 2001 [36]. This superconductor, developed at CERN with IASS [37] in collaboration with the Italian company Columbus [38] can be cooled using Helium gas (as opposed to liquid Helium), simplifying the demands on the cryogenic system. In addition, $MgB_2$ can function with a temperature margin of several degrees, which is a great advantage from the operation point of view.

Two first 10 m long $MgB_2$ cables have been successfully assembled and measured at CERN [37] (Figure 16). The cable element of 6.5 mm diameter, consisting of 18 $MgB_2$ strands and copper stabilizer, is capable of carrying a DC current of above 5 kA at 20 K. The current capability at 25 K is above 4000 A. As a next step, six of such elements will be assembled to reach the final objective of a 20 kA $MgB_2$ cable at 20 K. The measurements are done in a test station, recently built at CERN, which enables the measurement of up to 20 m long cables cooled by forced flow of Helium gas at any temperature in the range from 5 to 40 K.

The possible layout of the introduction of a super conducting solenoid in the present design of the ICARUS T600/T150 insulation/cooling system is sketched in Figure 17. The super-insulated helium cryostat, containing the SC coils, would be located in the narrow cooled interstice between the detector Aluminum cold body and the Nitrogen cooling system within the passive insulation container at an average temperature of about 87 K. From the construction point of view this layout has the advantage of minimizing the interference with the detector inputs/outputs, which are all already located on the top-side of the cold vessels.

In the case of the Magnesium Diboride ($MgB_2$) SC cable, refrigeration of the Helium coils (15-20 K) can be realized with the help of an additional commercial cryogenic cooler and with standard super-insulation layers ensuring a relative acceptable additional cryogenic loss towards the main vessel and the Nitrogen cooling system at the LAr temperature (87 K). If instead standard Niobium-Titanium superconducting cable will be used, currently available liquid Helium can be provided externally in relatively small quantities.

Preliminary calculations (Figure 18), assuming that all material within the coils are non magnetic, demonstrate that an appropriate vertical magnetic field ranging from 0.7 to 1.0 Tesla can be obtained in the T150 with a 6 m tall coil (exceeding the detector height by 1 m above and 1 m below) surrounding each of the two semi-modules separately. A current of 10-15 kA and ~100 turns/m are also required. The coil would then require ~29 km of cables conductor with a typical cross section of ~1-1.5 cm$^2$ and a total mass of ~26 t. These design parameters are not very different from those of the much larger and successfully operated ATLAS magnets. A similar coils layout around the T600 will ensure a slightly less uniform magnetic field, due to the longer detector length.

The estimated cost of the bare MgB$_2$ cable is according to Columbus estimates [38] of the order of 1-2 Euro/kA/m; in the T600 case the cost of the cable would be around 1 MEuros, therefore comparable to the cost of the LAr liquid supply for the indicated volumes. At present the stray magnetic field is allowed to extend over the whole volume around the detectors. This effect could be substantially reduced away from the detector with the help of some iron shield in the returning path.

## *2.7 Electronics.*

The ICARUS T600 electronics was successfully designed starting from an analogue low noise front-end amplifier followed by a multiplexed 10-bit AD converter and by a digital VME module that performs local storage and data compression [28], [29], [39]. The overall architecture, based on VME standard, is entirely appropriate for the experiment, nevertheless possible improvements are now conceivable, taking advantage of new more performing and compact electronic devices, for instance adopting a modern switched I/O and the parallelization of data flows.

An original feature of this layout is the housing of the electronics read-out channels onto the T150 flanges (Figure 19). Flanges are an original design of INFN already successfully used in the T600. In T150 we propose an evolution of the design that transforms the outer part of the feed-throughs' in a backplane for the front-end electronics. This design is already being tested on the small ICARINO prototype at the Legnaro National Laboratory (LNL) of INFN (Figure 20). A further R&D cooperative program could be initiated within the same ICARINO project, in order to explore the feasibility and performances of a cold, more compact electronics. This would imply improvements of the signal to noise ratio, shortening the length of signal cables, and a lower power consumption.

Performance, in terms of throughput of the read-out system, can be improved adopting a modern switched I/O, as PCI Express standard for instance, allowing the parallelization of the data flows. The use of low cost optical Gigabit/s serial links is under investigation, as well as implementing the White Rabbit (WR) serial high-speed connection.

## *2.8 Recombination phenomena in LAr.*

Saturation phenomena occur for heavily ionizing particles. None of the electron recombination theories developed so far is fully successful in describing all the experimental data in liquid argon. Nevertheless, they provide the basis for its understanding and for all phenomenological approaches.

The recombination effect has been already addressed by us [40] and it is a very important process, which needs to be further studied. There are several minor additives to the LAr, which modify this effect, transforming some of the produced light into additional ionization, which could be useful also for the future LBNE program.

The nonlinear detector response may degrade the particle identification capability of the LAr-TPC. A solution consists in introducing photo-sensitive dopants able to convert part of the scintillation light into additional free electron-ion pairs, thus enhancing the linearity as a function of the deposited energy density and electric field.

In past experimental tests of the doping systems TMG as photo-sensitive dopant has been chosen, since TMG is not absorbed in the recirculation system and can be easily purified to an electron lifetime better than 10 μs. TMG has a large photo-absorption cross-section of 62 MBarn and an acceptable quantum efficiency. The performance of the detector is greatly improved and is remarkably stable in time. A signal increase of +25% to +220% was found for stopping powers in between 1.6 to 32 MeV/cm [41] (Figure 21).

Initial tests will be performed in controlled conditions with the T150 detecting different charged particles (e, μ, π, K and protons).

### 2.9    Light Detection system.

An alternative solution to the traditional PMTs for the detection of light [42] is the use of solid-state photo-detectors, such as the Silicon Photomultipliers (SiPM) [43]. The cost of SiPM is continuously decreasing, making these objects almost competitive in terms of cost/cm$^2$ compared to large area photomultipliers. Advantages of this new solution would be: higher reliability, low power consumption, no high voltage power supply, negligible thickness, and possibility to operate in magnetic field. There are still some aspects that need to be explored, in particular: 1) the behavior at low temperatures; 2) the coupling with a proper wavelength shifter (WLS); 3) the choice of the best configuration (series and/or parallel) taking into account the high number of pixels of the array and the need to have a single signal cable; 4) the study of the electronic amplification and/or signal shaping.

## 3    A future for the ICARUS experimental program.

### 3.1    ICARUS proposal for FNAL.

As already mentioned, both the T600 and the new T150 will benefit of the experience matured with the T600 exposure to the CNGS beam. It is proposed to move the ICARUS LAr-TPC T600 to FNAL in a location common to the short baseline neutrino beam about 700 m from the Booster Beam (BNB) and at the same time exposed to the off-axis neutrino flux from the NuMI beam line (Figure 22).

The T150 detector will be located at a shorter distance of about 150 ± 50 m from the BMB. Both detectors will be magnetized with about 1 Tesla field installed during the provisional move to CERN. The ICARUS detectors will complement the already approved MicroBooNE program located at about 470 m.

With a total LAr volume of 760 ton, the ICARUS T600 detector has a unique role since it is presently the largest operational LAr-TPC and it will remain so for several years to come. It represents the status of the art and it marks a milestone in the practical realization of any future larger scale LAr detector. The T150 in a location at about 150 from the Booster is to be added.

Several other aspects of the detector will be revised during the provisional move to CERN, aiming at improving their performance. In particular, specific R&D will be carried out on cryogenics and LAr purity, readout electronics, electron-ion recombination phenomena, scintillation light detection as well as the detector calibration at test beams. This vigorous R&D program will be extremely beneficial also in view of the long-term implementation of the LAr-TPC technique for the LBNE multi-kton detectors.

### *3.2 The status of the search for sterile neutrinos.*

There are a number of *"anomalies"* which, if experimentally confirmed, could be hinting at the presence of additional, larger squared mass differences in the framework of neutrinos with mixing or of other effects. Two distinct classes of "anomalies" pointing at additional Physics beyond the Standard Model in the neutrino sector have been reported, namely a) the apparent reduction in the $\nu_e$ low energy neutrinos from nuclear reactors [44] and from the signal from Mega-Curie sources in the Gallium experiments [45], [46] originally designed to detect solar neutrino deficit, and b) evidence for an electron excess signal in interactions coming from neutrinos from particle accelerators [18], [19].

The class a) of phenomena hints to a significant fast disappearance rate in the initial $\nu_e$ production and the class b) predicts an anomalous $\nu_\mu \rightarrow \nu_e$ oscillation with similar, large $\Delta m_{new}^2$ values, much greater than the ones of the current three-neutrino mixing model. These experiments may all point out the possible existence of at least one fourth non standard neutrino state, driving neutrino oscillations at a small distance, with typically $\Delta m_{new}^2 \geq 0.5$ eV$^2$.

Therefore the possibility of short-baseline neutrino oscillations due to the existence of one or more additional neutrinos at the eV mass scale is a hot topic in current neutrino physics. Recent results [17] from the ICARUS experiment with the large mass T600 detector exposed to the CNGS neutrino beam at the LNGS laboratories, have considerably increased the evidence on the so far preferred 3+1 alternative for $\nu_e$ and $\nu_\mu$ appearance and disappearance in short-baseline experiments, given by the general formula:

$$P_{\nu_\alpha \rightarrow \nu_\beta} = \delta_{\alpha\beta} - 4|U_{\alpha 4}|^2 \left(\delta_{\alpha\beta} - |U_{\beta 4}|^2\right)\sin^2\left(\frac{\Delta m_{41}^2 L}{4E}\right).$$

Moreover the ICARUS experiment has demonstrated that the MiniBooNE low-energy anomaly (for its 3 lowest energy bins) is incompatible with ν oscillations and probably due to other (instrumental) effects. This result has been corroborated by the OPERA measurement with the CNGS beam [20].

The previously mentioned observed neutrino anomalies have to be accommodated in a global phenomenological framework that includes the present neutrino oscillation data in order to obtain a coherent description. In particular the relevant experimental results have been recently considered under the 3+1 neutrino model assumption [47]:

- The $\nu_\mu \to \nu_e$ appearance data of the LSND, MiniBooNE, BNL-E776, KARMEN, NOMAD, ICARUS and OPERA experiments;
- The $\nu_e$ disappearance data of the Reactor and Gallium anomalies;
- The constraints on $\nu_\mu$ disappearance obtained from the data of the CDHSW experiment, from the analysis of the data of atmospheric neutrino oscillation experiments, from the analysis of the MINOS neutral-current data and from the analysis of the SciBooNE-MiniBooNE neutrino and antineutrino data.

Global fits of $\sin^2(\theta_{\mu e})$ in appearance and of $\sin^2(\theta_{ee})$ and $\sin^2(\theta_{\mu\mu})$ in disappearance have been performed resulting in a preferred well defined common region $0.82 < \Delta m^2_{41} < 2.19$ eV$^2$ in agreement also with the expectations of cosmological results. A major contribution to this important result is provided by the ICARUS measurements, which allows to exclude most of the other ($\Delta m^2$, $\sin^2(2\theta)$) parameter area. The $\Delta m^2_{41} \approx 6$ eV$^2$ region is not allowed mainly because of old BNL-E776 data and cosmological exclusion. The best fit parameters from the global analysis from Ref. [47] ($\Delta m^2 = 1.5$ eV$^2$ and $\sin^2(2\theta) = 0.0015$) have been used.

The MiniBooNE experiment has been largely inconclusive at clarifying the LSND anomaly,: the crucial indication in favor of new short-baseline experimentation at accelerator is still given by the old and so far unexplained LSND result.

The present proposal aims at solving the problem of the LSND-like "anomalies" in the neutrino and antineutrino sectors and the related and possible existence of additional sterile neutrino flavors, searching for electron neutrino appearance in the almost pure, intense muon-neutrino beam at the FNAL-Booster, with a typical L/E range around 1 Km/GeV.

### *3.3 Sterile neutrino search at the Booster beam.*

The Booster Neutrino Beam line (BNB) exploits a 8 GeV primary proton beam, a beryllium target and a single horn focusing to provide a neutrino beam peaked around 0.8 GeV. The meson decay path is 50 m. The line develops horizontally a few meters below ground level. The BNB neutrino facility provides a high purity $\nu_\mu$ beam peaked around ~0.5 GeV, with a $\nu_e$ contamination at ~0.5 % level. Usually ~$4 \times 10^{12}$ protons are delivered each 0.9 s to the beryllium target in 1.6 μs long spills. About $2.2 \times 10^{20}$ pot (protons on target) are routinely delivered per year of data taking. The running of the BNB facility in antineutrino mode is also foreseen, although with lower beam intensity and larger neutrino contamination. Both the already approved MicroBooNE and ICARUS are LAr-TPC detectors.

The background due to neutral current events and to misidentified $\nu_\mu$ CC events is negligible with respect to the intrinsic νe component of the beam, because of the excellent imaging and calorimetric capability of the LAr-TPC detectors, which provides ~90% electron identification efficiency with a $\pi^0$ mis-interpretation probability of ~0.1% [4], [16], [17].

The integrated neutrino charged current event rates, for an exposure of three years ($6.6 \times 10^{20}$ pot) with positive focusing and five years ($11 \times 10^{20}$ pot) for negative focusing, are reported in Table 1 calculated for 150 m and 700 m from the target. These values can be approximately extrapolated from the 100 [48] and the 550 m [49] point as a function of the inverse of the distance squared. The corresponding energy spectra and composition at the two distances are shown in Figure 23, Figure 24 and Figure 25.

**Table 1.** Neutrino CC event spectra and composition in the Near (100 t fiducial mass at 150 m from target) and Far ((430 t fiducial mass at 700 m from target) detectors.

| Booster ν beam | Positive focus ($6.6\ 10^{20}$ pot) | | Negative focus ($11\ 10^{20}$ pot) | |
|---|---|---|---|---|
| Distance from target | 700 m | 150 m | 700 m | 150 m |
| E < 5GeV | | | | |
| $\nu_\mu$ | $3.87\ 10^5$ (100%) | $2.05\ 10^6$ (100%) | $5.12\ 10^4$ (90.2%) | $2.62\ 10^5$ (89.8%) |
| anti-$\nu_\mu$ | $2.79\ 10^3$ (0.72%) | $1.63\ 10^4$ (0.80%) | $5.69\ 10^4$ (100%) | $2.91\ 10^5$ (100%) |
| $\nu_e$ | $2.56\ 10^3$ (0.66%) | $1.47\ 10^4$ (0.72%) | $8.17\ 10^2$ (1.43%) | $4.12\ 10^3$ (1.41%) |
| anti-$\nu_e$ | 73 (0.02%) | $4.08\ 10^2$ (0.02%) | $3.12\ 10^2$ (0.38%) | $1.78\ 10^3$ (0.47%) |
| E < 2GeV | | | | |
| $\nu_\mu$ | $3.62\ 10^5$ (93.4%) | $1.92\ 10^6$ (93.8%) | $4.37\ 10^4$ (76.9%) | $2.42\ 10^5$ (83%) |
| anti-$\nu_\mu$ | $2.56\ 10^3$ (0.67%) | $1.53\ 10^4$ (0.74%) | $5.54\ 10^4$ (97.5%) | $2.90\ 10^5$ (99.5%) |
| $\nu_e$ | $1.90\ 10^3$ (0.49%) | $1.15\ 10^4$ (0.56%) | $5.72\ 10^2$ (1.01%) | $3.38\ 10^3$ (1.16%) |
| anti-$\nu_e$ | 44 (0.01%) | 250 (0.01%) | $2.36\ 10^2$ (0.41%) | $1.59\ 10^3$ (0.55%) |

As many as $\sim 4 \times 10^5$ $\nu_\mu$ CC events would be collected at the far position in positive focusing mode. The anti-ν CC rate in the negative focused beam is expected to be about a factor ten lower, because of the reduced negative meson production in target and the smaller anti-ν cross section. Indeed, in anti-neutrino mode the contribution of ν CC events is similar to the one of anti-ν CC events, resulting in a ~$10^5$ anti-$\nu_\mu$ + $\nu_\mu$ CC collected event sample at the far position. Moreover the corresponding νe beam component is three time larger than the anti-$\nu_e$ one, the total rising to ~1.4% of the anti-$\nu_\mu$ focused component below 2 GeV. As a result, the background for anti-neutrino oscillation search is three times larger that in the positive focusing beam, unless $\nu_e$ and

anti-$\nu_e$ events can be identified and separately measured with the introduction of the magnetic field in the LAr-TPC.

Possible background to $\nu_e$ and anti-$\nu_e$ due to cosmic ray photons entering the detector without accompanying charged particles has been investigated. A rough estimation of the rates foresees that ~0.3 such photon with energy greater than 200 MeV impinges on the detector within the 1 ms drift time. Such a background can be easily removed since:

i) As for $\pi^0$ rejection, the showers initiated by photons can be identified through the specific ionization in the first centimeters. Few residual singly ionizing events are due to Compton scattering, which in this energy range occurs with a probability of about $3\,10^{-2}$ with respect to pair production.

ii) It peaks in the downward vertical direction while genuine events point to the beam direction. It is possible to reject to about $10^{-3}$ cosmic photon with negligible reduction of efficiency for $\nu_e$ CC.

iii) The $\gamma$ interaction length in LAr is $\lambda_R = 14$ cm, and the length of not instrumented LAr crossed by $\gamma$ close to beam direction exceeds 1 m, reducing the background by a factor $\approx 0.5 \times 10^{-3}$.

iv) The $\nu_e$ event is producing also at least one hadron (proton, neutron, nuclear star if quasi elastic, plus a $\pi$ if inelastic), while the Compton $\gamma$ ray is recoil-less.

The total amount of observed ionization in the immediate vicinity of the vertex can signal the presence of a $\nu$-related event. The presence of a charged $\pi$ or a proton is an unambiguous signature, which cannot be missed. The presence of a neutron is more difficult to be detected since $\lambda_I = 80$ cm, although a sizeable local neutron star is expected in about 90% of the events.

In conclusion, the rejection capability of the CR gamma events is about $4 \times 10^{-8}$. Since the occurrence of these events is about one every ten beam pulses, the resulting background is negligible when compared to the expected $\nu_e$ CC (~$6 \times 10^{-6}$ per spill) for negative polarity beam in the far detector. Additional rejection can also be obtained from timing information provided by the scintillation light signals.

### *3.4 ICARUS LAr-TPC's installation at FNAL.*

Two new experimental halls have to be built along the neutrino beam line, to host the T600 and T150 detectors in a few meters deep pits, in order to match the beam position. The LAr vessels and the thermal shields could be inserted from the roof. The required pit area for the installation of the T600 LAr-TPC modules is 29 m (length) × 20 m (width). A 20 m height from the pit floor is required. A similar layout of Near site hall is required with a pit area for the installation of the T150 of 25 m (length) × 15 m (width).

It is expected that the three cryostats, containing the fully equipped inner detectors, will be transported to FNAL the after overhauling of the T600 and the construction of the T150. The various ancillary components transported from LNGS and CERN will be organized in containers, which should be stored in an appropriate area, before the final installation in the detector areas.

The T600 installation in the Far detector site is shown in Figure 26. The pumps for liquid argon recirculation and the relative filtering units will be positioned in the 3 m wide area left in front upstream of the T600 cryostat together with the services for the vessels filling and emptying. The Stirling cryo-coolers (floor level), the liquid nitrogen storage tank (top level) and other cryogenic equipment should be positioned on the T600 side (blue area shown as cryo-vessels in Figure 26), into an appropriate supporting structure. The storage tanks and other services will be positioned at a height of about at least 5 m from the floor (ground level) to allow for the liquid nitrogen circulation in pure passive mode. Additional storage tanks (e.g. for Liquid Argon during the filling and emptying phases) will be located outside the building, close to the cryo-vessels area and connected to cryogenic transfer lines. A shielding layer capable to absorb the soft component of the cosmic ray background, which has to be foreseen above the detector, will be inserted at the level of the electronic racks.

The three cryostats, containing the fully equipped inner detectors, will be transported to FNAL after overhauling of the T600 and the construction of the T150. The various ancillary components transported from LNGS and CERN will be organized in containers, which should be stored in an appropriate area, before the final installation in the detector areas.

The T150 installation, at the Near location, should be organized following a similar layout. As for T600, the storage tanks and other services should be positioned at a height of about at least 5 m from the floor (ground level), to allow for the liquid nitrogen circulation in pure passive mode and the Stirling liquefiers should be installed at ground level.

## 3.5   Trigger and DAQ.

The trigger system of both the T600 and the T150 detectors will exploit the coincidence of the prompt signals from the scintillation light in the LAr-TPC with the proton spill extraction of the Booster within a gate of about 2 µs.

At the nominal BNB intensity of $4 \times 10^{12}$ pot / spill, ~1 neutrino interaction with vertex in the LAr-TPC's every 400 spills is expected to trigger the T600 at the far position. A similar trigger rate will come from beam associated through-going muons while the cosmic rays events will produce a trigger every 30 spills.

Due to the reduced distance from target, the corresponding beam associated trigger rates for the T150 at the near position are increased to 1 neutrino interaction / 200 spills and 1 beam associated muon / 100 spills. The trigger rate from cosmic rays will be reduced to 1/120 spills due to the smaller detector mass.

At the standard BNB repetition rate of 4 Hz, about 1 event every 6 s in the T600 at the Far position and 1 event every ~10 s in the T150 at the Near location are expected. The trigger rates would be well within the T600 DAQ throughput already achieved for the CNGS data taking at LNGS (~1 Hz) even for the maximum rate available after Booster upgrade (~15 Hz).

The present architecture of the ICARUS T600 DAQ system allows handling each readout chamber (made of 24 readout units), with a maximum throughput of ~50 MB/s, safely exploiting the data link at half of the available bandwidth. All the readout units can work autonomously, pushing their own data to the receiving workstation. Segmentation and parallelization of the data stream (e.g. 12 readout units per builder unit) allow reaching a building rate > 1 Hz on the whole T600, largely adequate to match the expected trigger rate. Additional detector segmentation would allow even higher throughput rates.

### 3.6 *Complementing MicroBooNE with ICARUS at BNB.*

The MicroBooNE detector foreseen during 2014 at FNAL at the distance of 470 m from the target will be the first LAr-TPC detector specifically designed for physics operation [25]. The MiniBooNE experiment has apparently not entirely confirmed the presence of the LNSD signal above 480 MeV neutrino energies, but it has introduced a new low energy effect at lower energies. The ICARUS [17] and OPERA [20] experiments have already concluded on the instrumental nature of this new MiniBooNE low energy anomaly, but at a much larger $L/E_\nu$. Their actual origin will be studied by the MicroBooNE detector, which will for instance unambiguously identify electrons from single photons.

A further search for the LSND-like anomaly with neutrinos and a single MicroBooNE detector is however strongly limited by the presence of beam related uncertainties. The uncertainty of the $\nu_e$ intrinsic background, in the configuration with a single detector experiment is ~15% mainly due to systematic uncertainties on kaon production at the relevant energies, estimated to be at the 30% level [26]. Examples of $\nu_e$ event spectra in presence of $\nu_\mu \rightarrow \nu_e$ oscillations (positive polarity beam) are compared with the best fit parameters $\Delta m^2 = 1.5$ eV$^2$ and $\sin^2(2\theta) = 0.0015$ from the global analysis of Ref. [47] retained as an example (see Figure 27 and Table 2)

A proposal for a dual baseline experiment has been initially presented at CERN as early as 2009 [24]. The addition of ICARUS now proposed for FNAL is an important complement to MicroBooNE, permitting a search for spectral differences of electron like specific signatures *in two identical detectors but at two different neutrino distances*, at the Far and the Near locations. In absence of oscillations, after some beam related small corrections required by the beam line geometry and focusing system, the two energy spectra should be a precise copy of each other, independently of the specific experimental event signatures and without any need of Monte Carlo comparisons. Therefore an exact, observed proportionality between the two $\nu_e$ spectra directly implies the absence of neutrino oscillations over the measured interval of L/E: any resulting $\nu_e$ difference between the two locations, if observed, must be inevitably attributed to the time evolution of the neutrino species.

Our proposed experiment, collecting a large amount of data both with neutrino and antineutrino focusing, should be able to give a likely definitive answer to the 4 following queries:
- the LSND + MiniBooNe both antineutrino and neutrino $\nu_\mu \rightarrow \nu_e$ oscillation anomalies;
- The Gallex + Reactor oscillatory disappearance of the initial $\nu_e$ signal, both for neutrino and antineutrinos
- an oscillatory disappearance maybe present in the $\nu_\mu$ signal, so far unknown.
- Accurate comparison between neutrino and antineutrino related oscillatory anomalies, maybe due to CPT violation.

In the two detector configuration of this proposal, the sensitivity to sterile neutrino oscillation search has been computed according to the particle identification efficiency described above and assuming a 4% uncorrelated systematic uncertainty in the prediction of "Far" to "Near" $\nu_e$ ratio. Indeed, only small kinematical corrections are needed in order to predict the spectra at the far position, due to the difference in solid angle and to the fact that the near detector sees an extended line source instead of a point source.

On the contrary, in the case of a single detector setup, the neutrino fluence at the detector has to be completely predicted by detailed MC calculation starting from the primary proton beam interaction in the target. Therefore, in a single detector experiment, the major systematic uncertainties come from the knowledge of particle production in the proton target, the focusing system and neutrino cross-sections. In the BNB, about one half of the electron neutrino background events are produced by Kaon decay, while the muon neutrinos are essentially produced by pions. Systematic uncertainties on kaon production at those energies have been quoted [26] at the 30% level, however the lack of experimental data does not allow for a precise estimation.

Moreover, electron and muon neutrino cross sections are slightly different in the BNB neutrino energy range, and their difference is also affected by poorly known nuclear effects. From the point of view of the neutrino event measurement and reconstruction, the use of a single detector will rely heavily on detailed MC simulation and reconstruction in term of detector response signal efficiencies and associated backgrounds. All these detector related effects that consequently introduce large systematic uncertainties are no longer present in a two-detector experiment. This cancellation is maximized in the case of two identical near and far detectors.

The effect of the oscillation at the near position on the sensitivity is visible in the high $\Delta m^2$ region, well above the island defining the allowed parameter range according to past experiments.

**Table 2.** Examples of $v_e$ appearance event rates at the far detector in presence of $v_\mu \to v_e$ oscillations (positive polarity beam). The best fit parameters from the global analysis of Ref. [47] ($\Delta m^2 = 1.5$ eV$^2$ and $\sin^2(2\theta) = 1.5 \times 10^{-3}$) have been used. An exposure of $6.6 \times 10^{20}$ pot has been assumed. The oscillation sensitivity is found to be poorly affected by far detector distance in the range 550-700 m.

|  | E < 5 GeV | E < 2 GeV |
|---|---|---|
| Far detector distance: 550m | | |
| Intrinsic $v_e$ | 4.19 10$^3$ | 3.11 10$^3$ |
| Signal + intrinsic $v_e$ | 4.80 10$^3$ (115%) | 3.71 10$^3$ (119%) |
| Far detector distance: 700m | | |
| Intrinsic $v_e$ | 2.52 10$^3$ | 1.86 10$^3$ |
| Signal + intrinsic $v_e$ | 2.92 10$^3$ (116%) | 2.24 10$^3$ (121%) |

Since the negative polarity beam is highly contaminated by the neutrino component, a reasonable sensitivity to the antineutrino oscillations can be achieved only by discriminating the two components in the measured events, with the help of a magnetic field. Otherwise, the high level of intrinsic electron neutrino background would spoil the measurement. In order to definitely clarify the origin of the anomalous LSND antineutrino oscillation signal, and in view of a possible CPT non conservation, only anti-$v_\mu \to$ anti-$v_e$ oscillations have to be considered in the negative beam polarity. Therefore, the sensitivity curves for negative polarity beam are obtained assuming oscillation of the muon anti-neutrino component only, and two hypothesis for the background, namely with and without discrimination of the $v_e$ and anti-$v_e$ contributions to the intrinsic background.

A sensitivity of $\sin^2(2\theta_{new}) > 3.1 \times 10^{-4}$ (for $\Delta m_{new}^2 > 1.0$ eV$^2$) at 90% C.L. is expected for the $v_\mu \to v_e$ transitions with three year exposure ($6.6 \times 10^{20}$ pot) at the Booster $v_\mu$ beam (Figure 28). The parameter space region still allowed by the LSND/MiniBooNE anomalies is fully covered even at 99% C.L.

In anti-neutrino focusing mode, nearly twice as much exposure ($11 \times 10^{20}$ pot) is required to let the allowed anomalies regions fall well within the reach of this proposal (Figure 29). A sensitivity of $\sin^2(2\theta_{new}) > 6 \times 10^{-4}$ (for $\Delta m_{new}^2 > 1.0$ eV$^2$) at 90% C.L. is expected for the anti-$v_\mu \to$ anti-$v_e$ transitions with a magnetized LAr-TPC. In absence of magnetic field, a 5 year run is not entirely sufficient to cover the allowed parameter space region even at 90% confidence level.

The persisting presence of disappearance anomalies in reactor neutrino experiments, in the Gallex and Sage calibration with sources and in other experiments, is an open challenge. As well known, these data may be well fitted by the 3 + 1 neutrino hypothesis, while the no-oscillation hypothesis is disfavored at 99.93% C.L. The experiment presently proposed is however intended to detect experimentally the oscillation pattern in the $\sin^2(2\theta) - \Delta m^2$ plane, beyond the so far unaccounted lack of $\nu_e$ events. Both the $\nu_e$ and $\nu_\mu$ disappearance signals may be searched for.

In Figure 30 the energy distributions of the electron neutrino events is shown for the "Far" (a) and "Near" (b) positions, together with a number of possible $\Delta m^2$ value in the region of $\Delta m^2 > 1$ eV$^2$ and $\sin^2(2\theta) \approx 0.16$. If confirmed, the existence of a new neutrino species carrying such a large mass will have an important role in the explanation of the existence of the Dark Mass in the Universe.

In Figure 31 the 90% confidence levels for the actual oscillation mechanism in the $\sin^2(2\theta) - \Delta m^2$ plane are shown with an integrated intensity corresponding to $6.6 \times 10^{20}$ pot. They are also compared with the "anomalies" from the combination of the published reactor neutrino experiments Gallex and Sage calibration sources experiments.

The disappearance signal in the same $\sin^2(2\theta) - \Delta m^2$ range may also be studied with the dominant $\nu_\mu$ and anti-$\nu_\mu$ signals. The $\nu_\mu$ and anti-$\nu_\mu$ spectral shapes, primarily due to pion decays, are significantly different in the "Near" and "Far" positions. In the energy range below 2 GeV, where the effect is expected, the relative differences amount to about 30% and they may be predicted to about 4%. This systematic uncertainty is larger than the statistical fluctuations expected for the huge number of collected events. However, the amount of $\nu_\mu$ depletion should be large enough for a significant test.

### 3.7 Neutrino interactions studies with the Off-Axis NUMI beam.

The NuMI beam-line is fed by 120 GeV protons with $4 \times 10^{13}$ protons per pulse. The secondary beam includes a double horn focusing system which allows for different variable energy configurations producing a neutrino beam directed downwards, towards the far MINOS detector, with a slope of ~50 mrad.

As described by the MiniBooNE and MicroBooNE Collaborations, neutrinos produced by the NuMI beam reach the far locations of the Booster beam line. These "off-axis" neutrinos are a valuable tool for the study of neutrino cross-sections and interaction topologies at energies of interest for the LBNE experiment.

Given the NuMI repetition rate (0.53 Hz) and its spill duration (8.6 μs), one additional trigger every 12 s is expected, mainly due to cosmic rays occurring in the coincidence gate. This increase in trigger rate will not affect the DAQ performance. About 1 neutrino event from NuMI every 150 s is foreseen.

The T600 will collect a large neutrino event statistics in the 0 to 3 GeV energy range with an enriched component of electron neutrinos (several %) from the dominant three body decay of secondary Kaons. A careful and detailed analysis of these events will be highly beneficial for the future LBNE LAr program, allowing to study very precisely detection efficiencies and kinematical cuts in all neutrino channels and event topologies.

A FLUKA [50] based Monte Carlo simulation of the NuMi beam line has been set up according to the available technical drawings [51] for the low energy beam configuration. The obtained neutrino fluxes have been compared with those published by the MINOS collaboration at the MINOS near detector position. Even if not all the geometry details were available and/or included in the simulation, our results agree with the MINOS ones within 20%, indicating that reliable prediction of neutrino rates at off-axis positions are possible.

In Figure 32 the neutrino rates at the present MiniBooNE location, ~105 mrad off-axis at ~880 m from the NuMI target, are shown. In the muon neutrino spectrum there is clear evidence of a very low energy component coming from pion decay and a peak at about 2 GeV due to kaon decay. The electron neutrino spectrum originates essentially from kaon decay and presents a broad peak covering the 0.5-2 GeV energy range.

**Table 3.** CC event rates from the NuMI beam at the far (700 m) position for 1 year exposure ($3 \times 10^{20}$ pot) with the T600. The electron neutrino component is about 4% of the total event rate.

| $\nu_\mu$ | $\bar{\nu}_\mu$ | $\nu_e$ | $\bar{\nu}_e$ |
|---|---|---|---|
| $1.8\ 10^5$ | $3.1\ 10^4$ | $7.9\ 10^3$ | $1.5\ 10^3$ |

The expected event rates are summarized in Table 3. Muon neutrino event rates are comparable with the ones from the Booster beam. However, the electron neutrino component is enhanced in the off-axis beam, with as much as 8000 events/year in the T600. This amount of data would allow a detailed evaluation of detection efficiency and background reduction at the energy of the second oscillation maximum in the LBNE expected signal.

# 4  Time schedule and infrastructures.

The ICARUS T600 decommissioning at LNGS Gran Sasso INFN Laboratories is smoothly proceeding before the expected overhauling at CERN. In parallel the R&D activities on the topics discussed in this proposal have already started at CERN, LNL (Legnaro) and LNGS INFN Laboratories, as well as the requested studies for the new cold vessels, passive insulation and the general infrastructures. The prototypes of the new and more compact read-out electronics will be soon tested with the ICARINO LAr-TPC test facility at Legnaro.

The foreseen schedule needed for the T600 LAr-TPC overhauling at CERN and the new T150 construction fits a 2 years time window, starting with the "green light" by the Funding Agencies (Figure 33). The two detectors are expected to be operational at FNAL within 2016, ready for data taking with Booster and NuMI neutrino beams.

The experiment should be installed in two experimental halls where the two pits (Far: area $29 \times 20$ m$^2$, height ~20 m), (Near: area $25 \times 20$ m$^2$, height ~15 m), realized according to the FNAL Standards and Rules, will be equipped by:

- Cranes: T600 (Far) with 50 t capacity, T150 (Near) whit 30 t capacity;
- Mains: T600 $2 \times$ trasfos 1 MVA each, T150 $2 \times$ trasfos 0.5 MVA each. Power supplies must be redundant. The general services (light, cooling, ventilation, heating, etc.) are not included in this estimate;
- Air conditioning, heating: no special requirements;
- Cooling: demineralized cooling water is needed in both Far and Near buildings.

Specific safety requirements:

- Ventilation: following FNAL safety instructions; a 2 flow rates system is recommended, the reduced flow rate always running, and the higher in service in case of alarm and the aspiration at ground level, the rejection outside the halls;
- Safety sensors: oxygen, smoke, temperature;
- Emergency lights;
- Cameras to control the ICARUS area;
- Audio Alarms.

A general Control Room should be implemented in the Far detector building as unique data receiving and control point for both T600 and T1500 detectors.

The data storage and computing power resources implies the use of FNAL facilities. As for LHC, a TIER 0 structure should be setup, with dedicated connection to the DAQ system. Long-term data storage should also be provided.

During the assembly of the experiment several persons of the Collaboration are expected to work on site: they need an adequate number of offices, access and use of mechanical and electrical workshops, and the use of an appropriate area for storage of their specific materials. A dedicated electronics workshop, equipped with testing instrumentation (oscilloscopes, signal generators, power supplies etc.) will be also required for possible interventions on the detector electronics.

# 5  Figures.

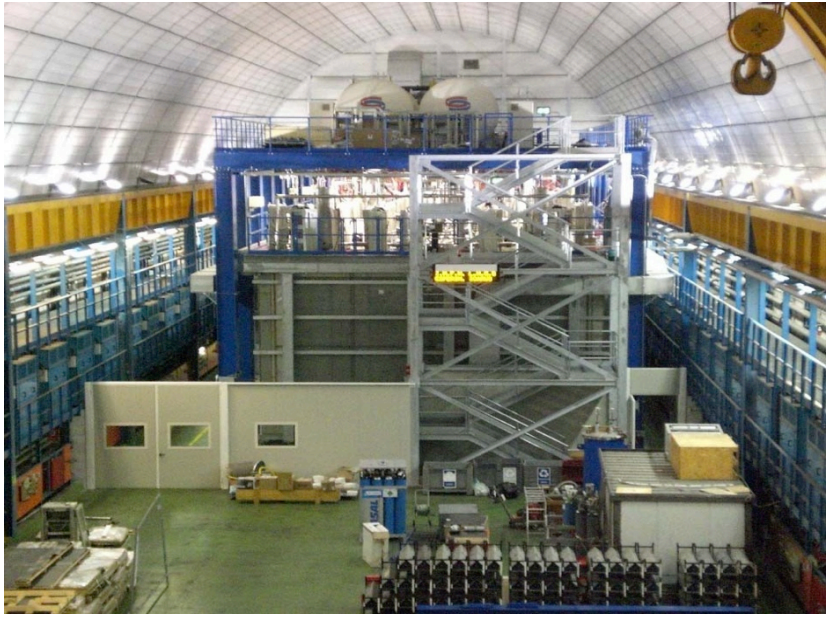

**Figure 1.** Picture of the ICARUS T600 installation in the Hall-B of the LNGS underground laboratory.

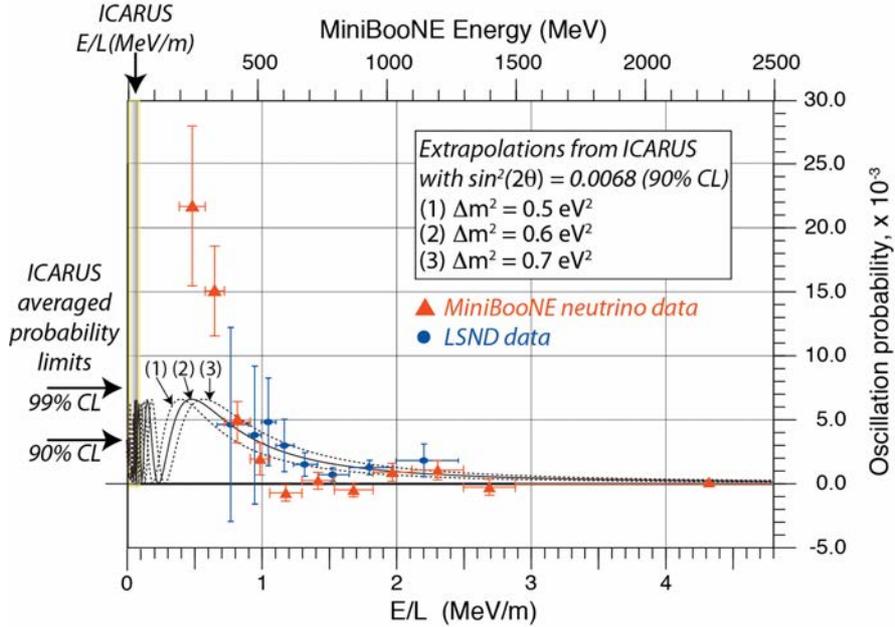

**Figure 2.** Oscillation probability limits obtained by the ICARUS experiment at LNGS with the CNGS neutrino beam [17] compared with corresponding data from neutrinos in MiniBooNE [19] as a function of the energy $E_\nu$. Figure 2 in Ref. [19] has been used in order to convert the observed number of excess events/MeV to their corresponding oscillation probabilities. In order to perform the conversion, the values $\sin^2(2\theta_{new}) = 0.2$ and $\Delta m^2_{41} = 0.1$ eV$^2$ from Figure 2 of Ref. [19] have been used. The resulting oscillation probability distribution for neutrino and for $E_\nu > 475$ MeV appears compatible with the absence of antineutrino LNSD effect. For the $200 < E^{QE} < 475$ MeV region (triangular red points) - below the sensitive $E_\nu/L$ region of LSND - the new MiniBooNE effect $\nu$ is widely incompatible with the averaged upper probability limit to anomalies from ICARUS [17] and OPERA [20] on $\sin^2(2\theta_{new})$ in their $E_\nu/L$ regions. An extrapolation from ICARUS to larger values of $E_\nu/L$ for two-neutrino oscillation parameters simultaneously compatible with LSND and other previous experiment is also shown as guidance.

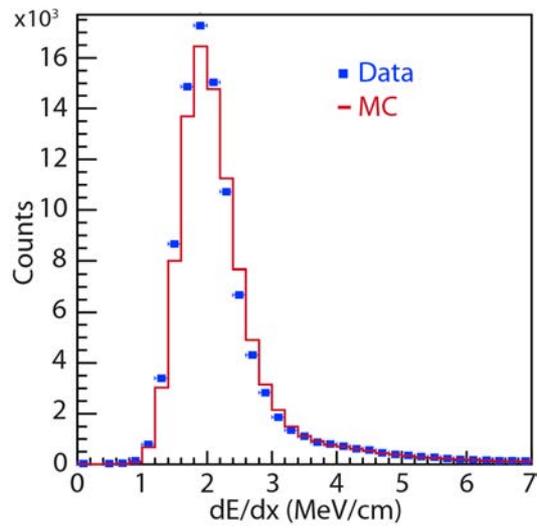

**Figure 3.** Energy deposition density distribution for muons in CNGS CC interactions, compared with Monte Carlo, normalized to the same number of entries. Each entry corresponds to the energy collected by a single wire.

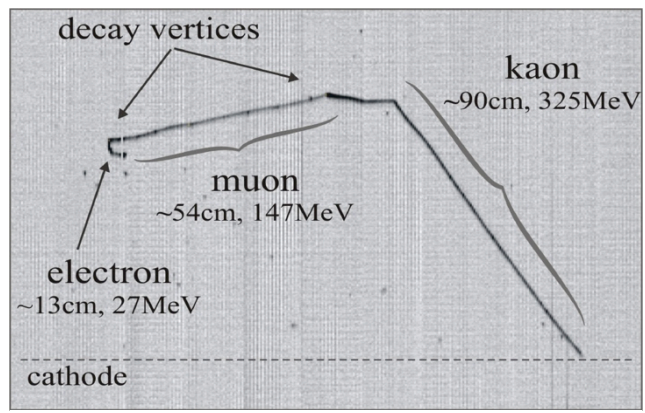

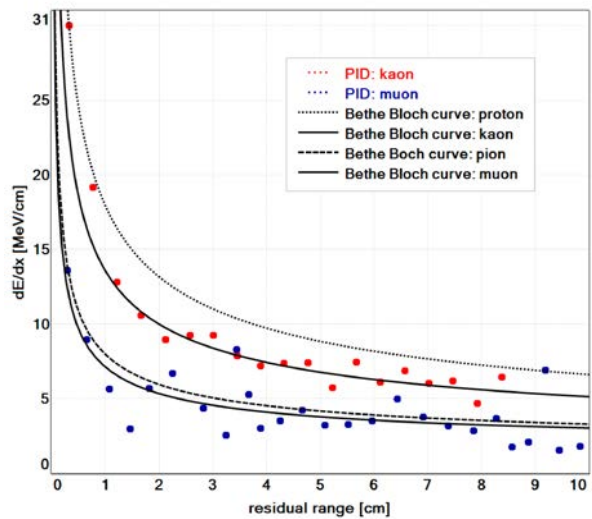

**Figure 4.** Example of kaon decay in a CNGS event (top) and corresponding PId patterns (bottom).

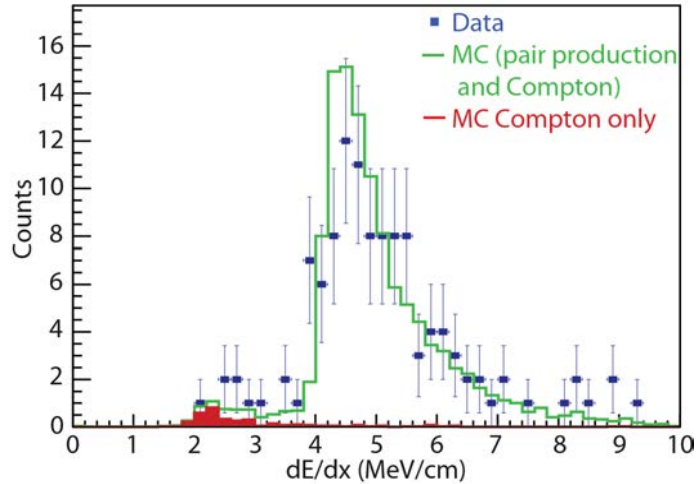

**Figure 5.** Average ionization in the first 8 wire hits for sub-GeV photons in the T600 data (full squares), compared to Monte Carlo expectations normalized to the same number of events. In MC case, the Compton contribution is shown also separately.

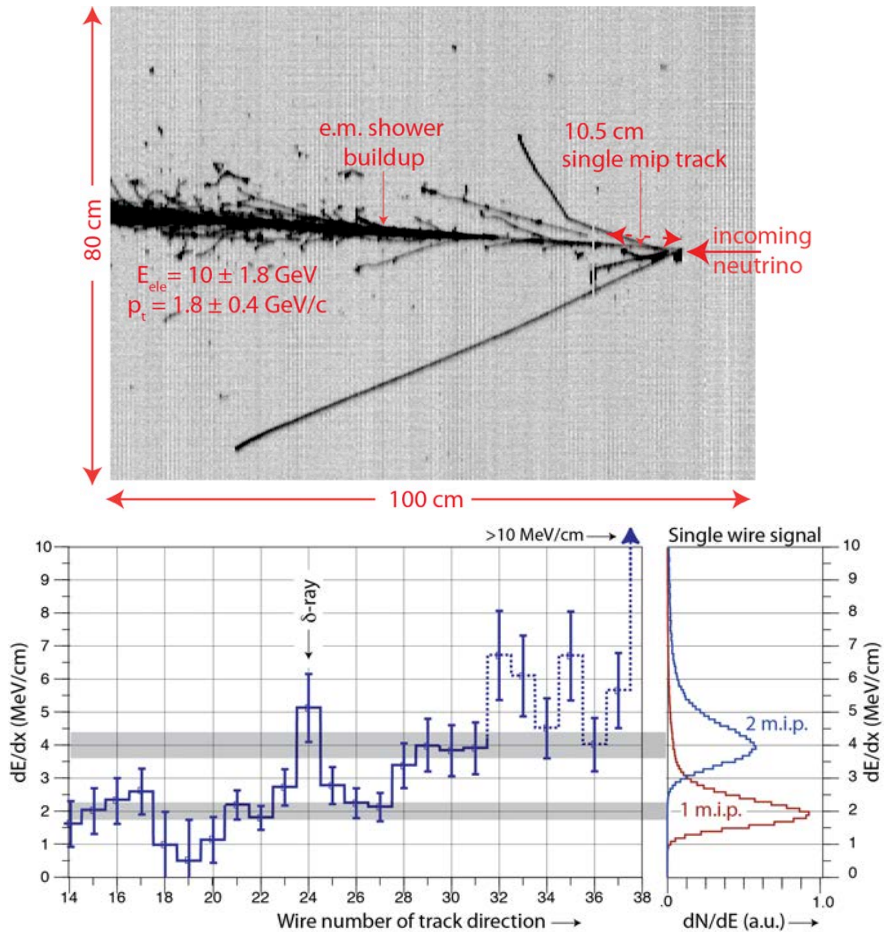

**Figure 6.** (Top) Experimental pictures of an identified electron neutrino event found in the CNGS data sample. The event has a total energy of $11.5 \pm 1.8$ GeV and an electron of $10 \pm 1.8$ GeV with a transverse momentum of $1.8 \pm 0.4$ GeV/c. (Bottom) The evolution of the dE/dx from a single track to an e.m. shower for the identified electron is shown along the individual wires in the region ($\geq 4.5$ cm from primary vertex) where the track is well separated from other tracks and heavily ionising nuclear prongs. As a reference, the expected dE/dx distribution for single and double minimum ionising tracks are also displayed.

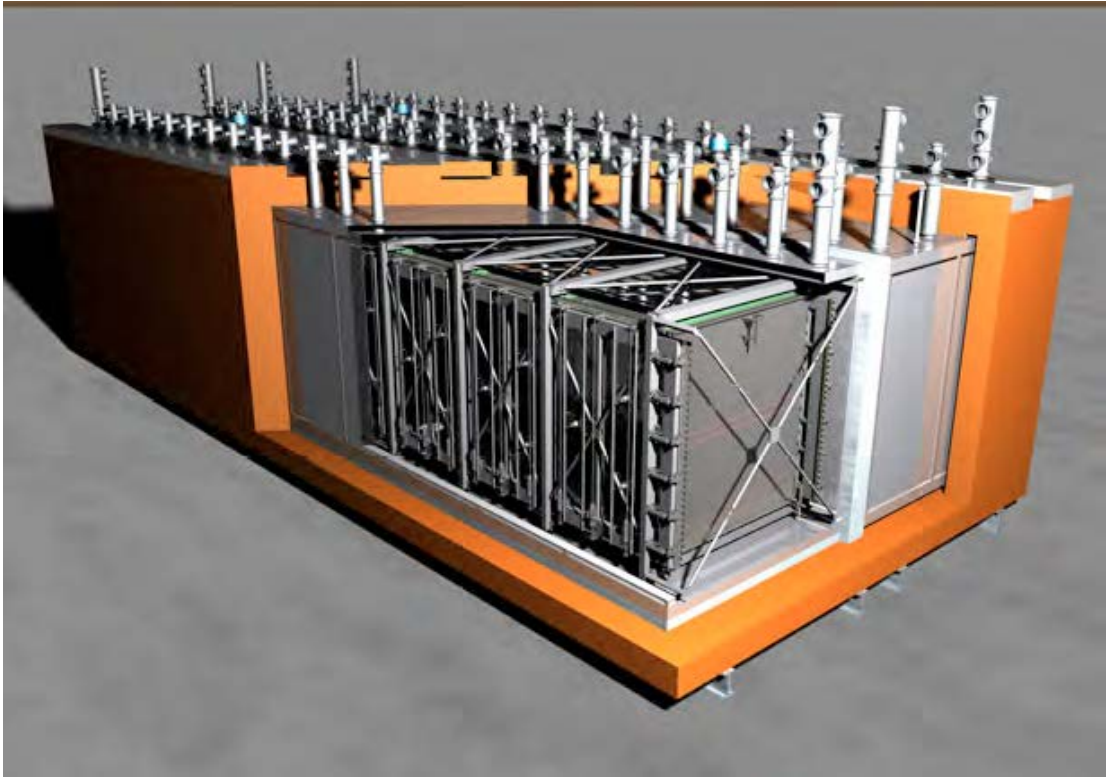

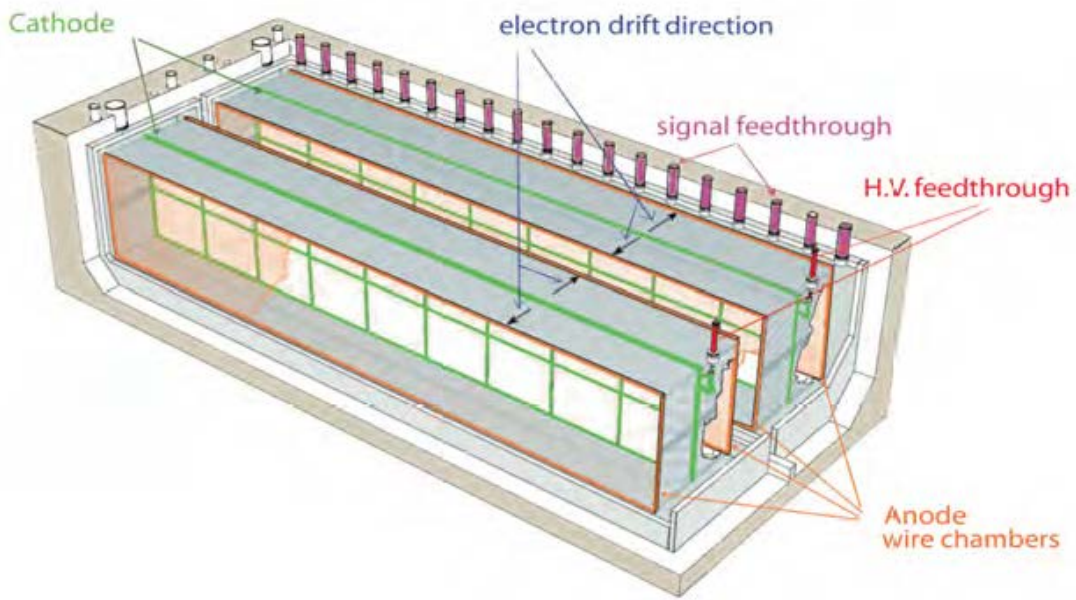

**Figure 7.** ICARUS T600 detector schematics, showing both modules and the common insulation surrounding the detector; inner structures and feed-through's are also shown. On the bottom a view of the detectors is presented with the wire chambers and the high voltage system (race-tracks and cathodes).

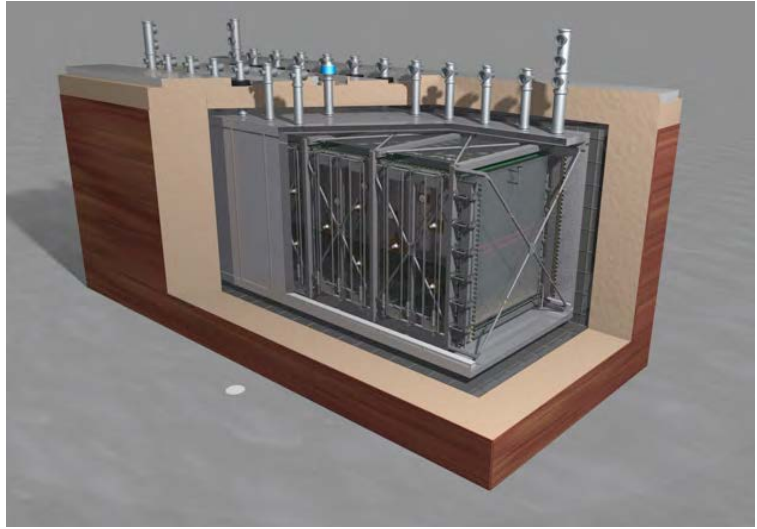

**Figure 8.** 3D view of the new ICARUS T150 detector.

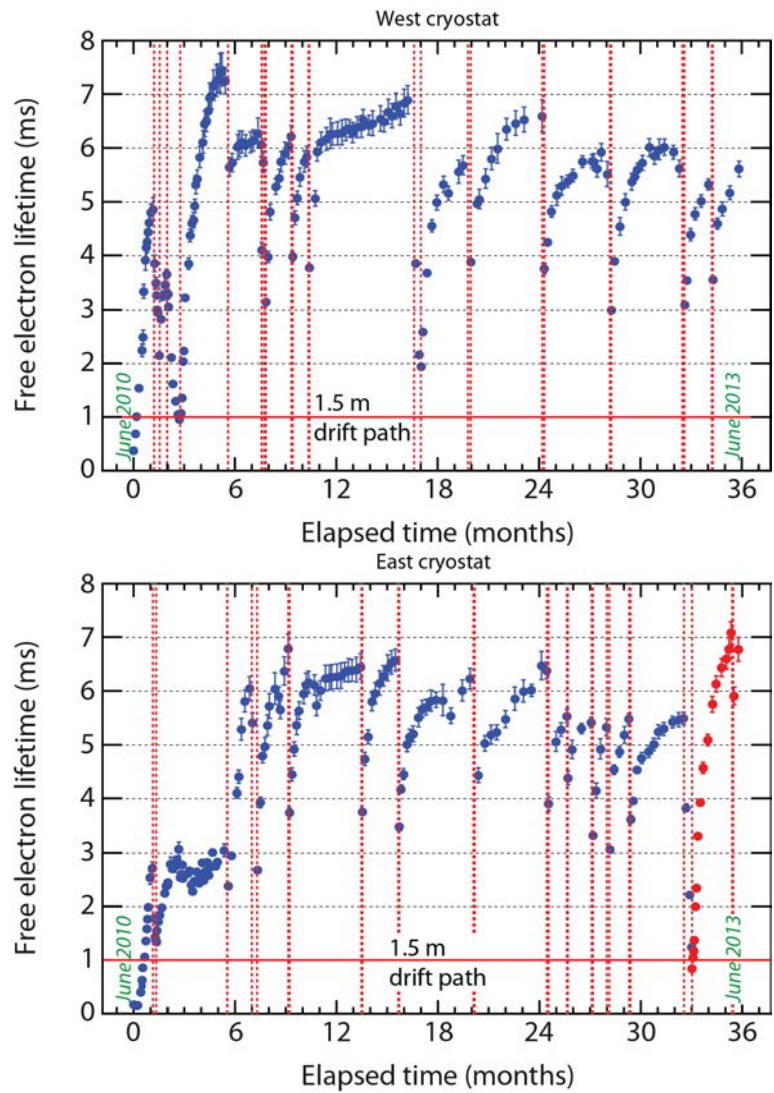

**Figure 9.** Electron lifetime evolution in West and East modules for the full T600 live-time at LNGS. LAr recirculation stops for maintenance are marked as dotted vertical lines. Values measured after the installation of a new recirculation pump with improved performance are marked in red.

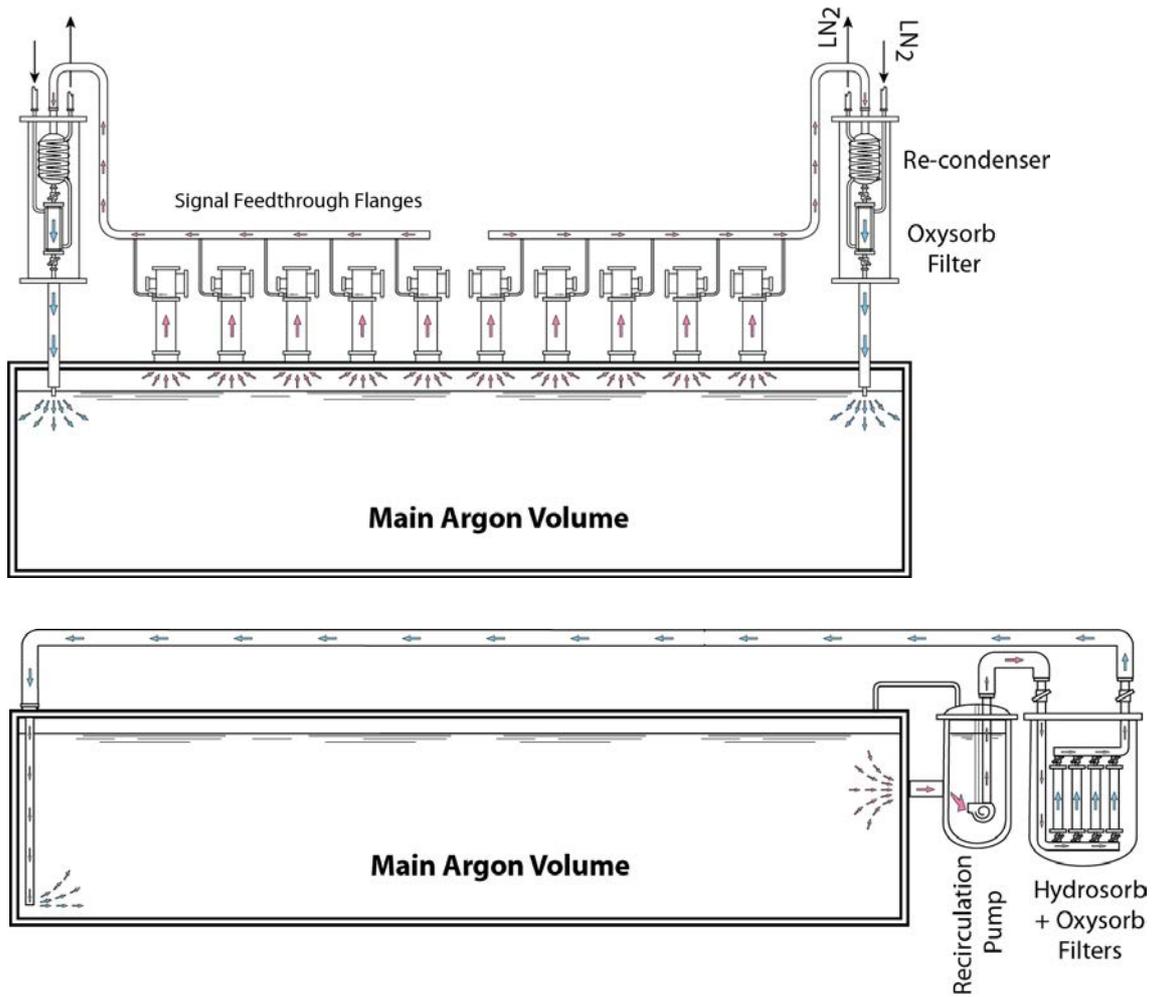

**Figure 10.** Schemes of the GAr (top) and LAr (bottom) recirculation / purification systems of the T600.

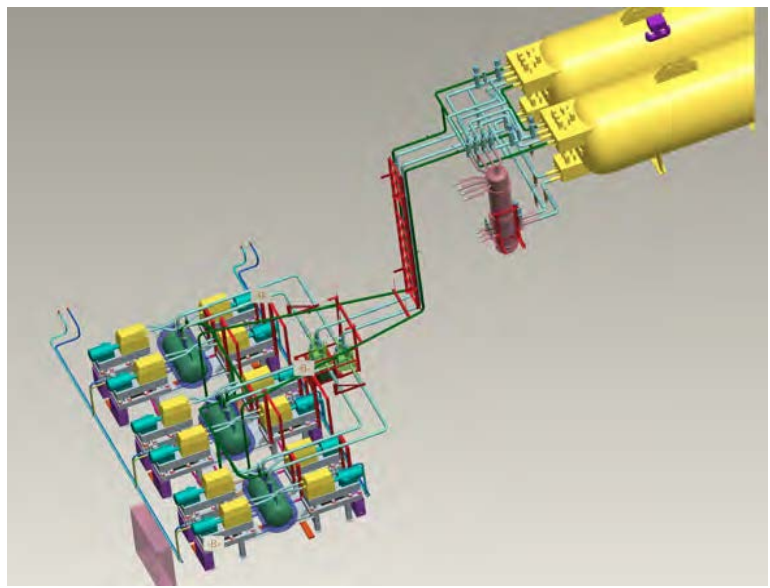

**Figure 11.** General view of the Nitrogen re-condensation system based on the Stirling cryo-coolers as in the LNGS installation.

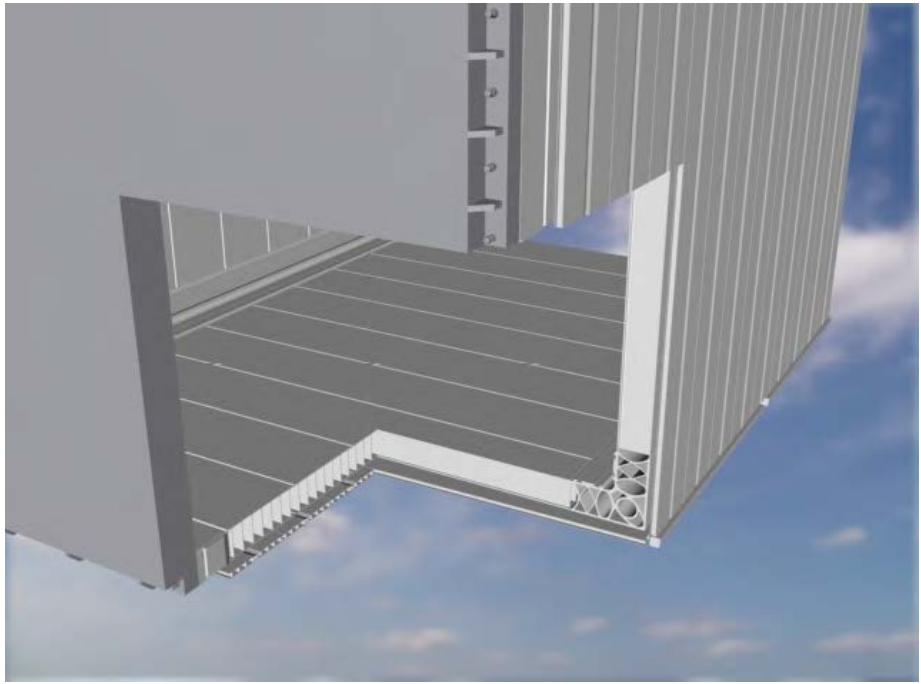

**Figure 12.** General view of a cross-section of the new aluminum vessels. The external cooling shield is also visible.

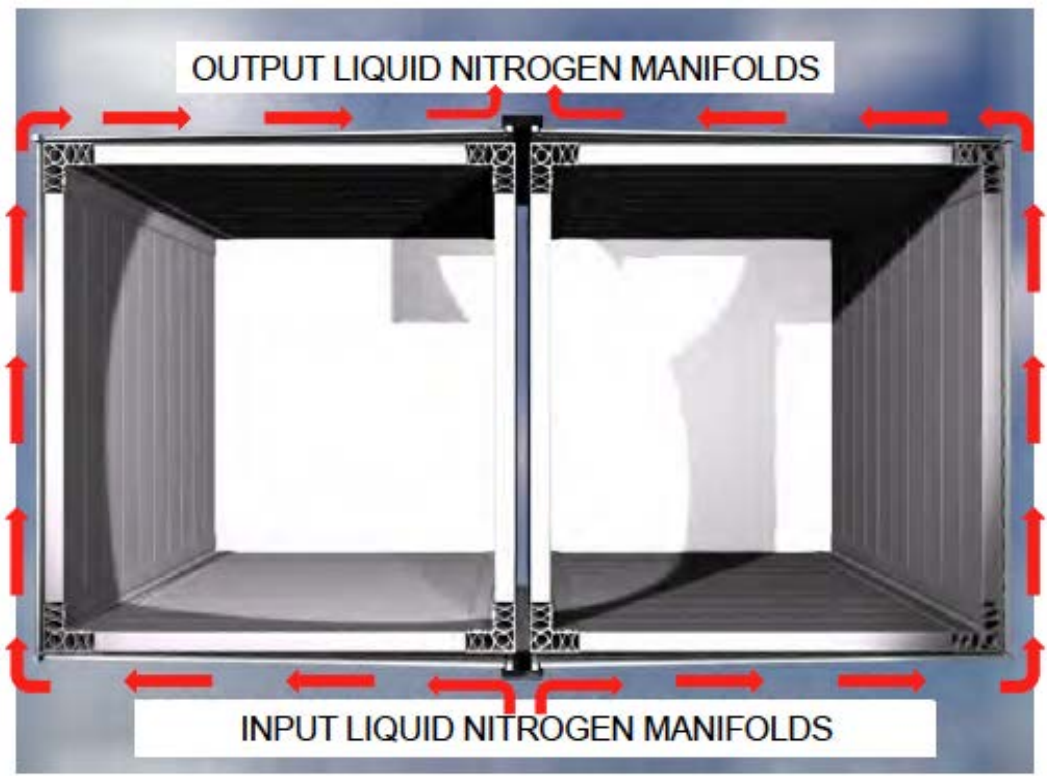

**Figure 13.** General view of the Nitrogen shields for the T600. Nitrogen flow is indicated with the red arrows.

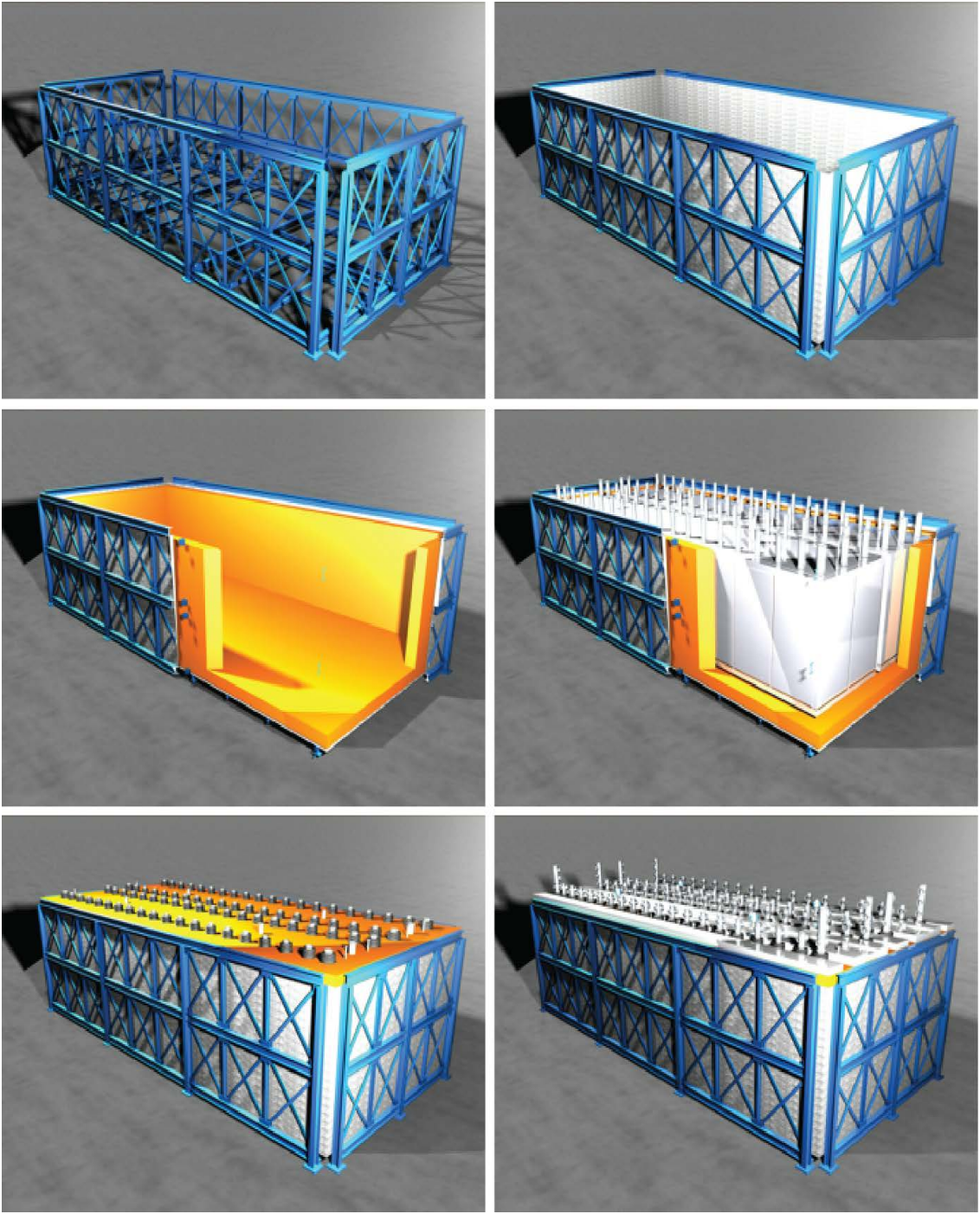

**Figure 14.** The new T600 insulation scheme. Top-left: warm vessel cage. Top-right: external skin. Center-left: insulation panels. Center-right: T600 cols vessels. Bottom-left: insulation top panel. Bottom-right: top flanges.

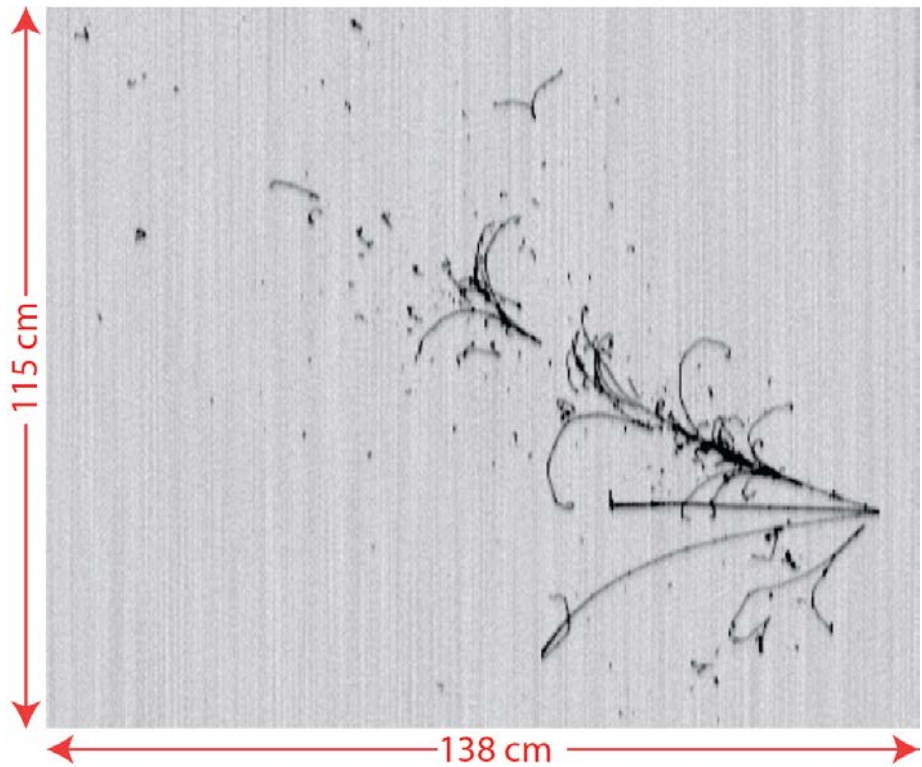

**Figure 15.** MC simulation of a 4 GeV CC $\nu_e$ event in a magnetized LAr-TPC as seen in collection view. The final state contains an electron, a $\pi^0$, a $\pi^+$ and a proton. The magnetic field is 1 Tesla, orthogonal to the picture.

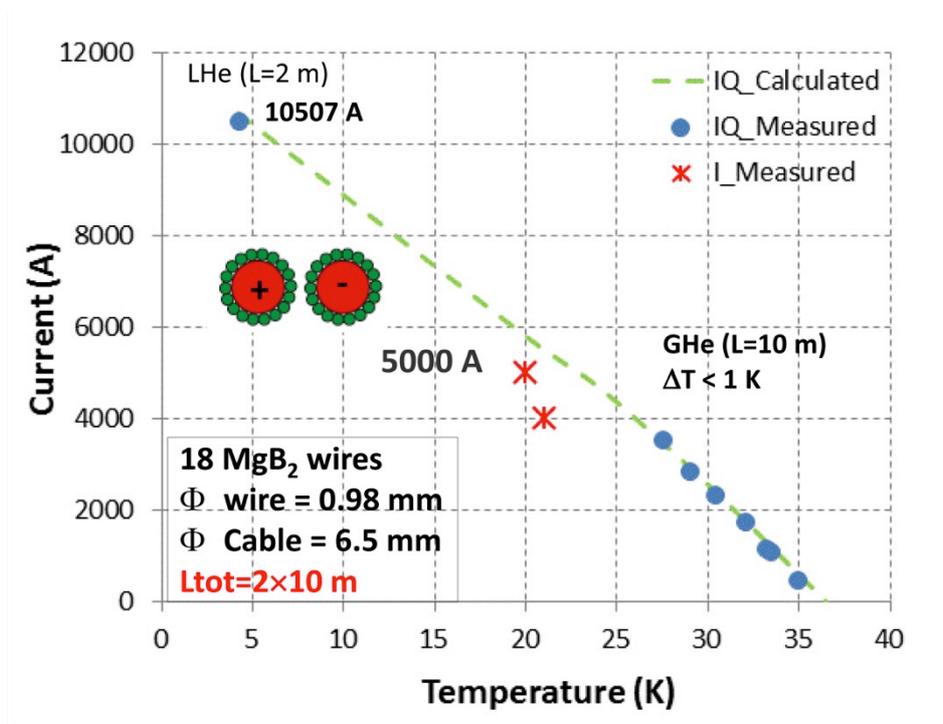

**Figure 16.** The first 10 m long $MgB_2$ cable recently successfully assembled and tested at CERN [37].

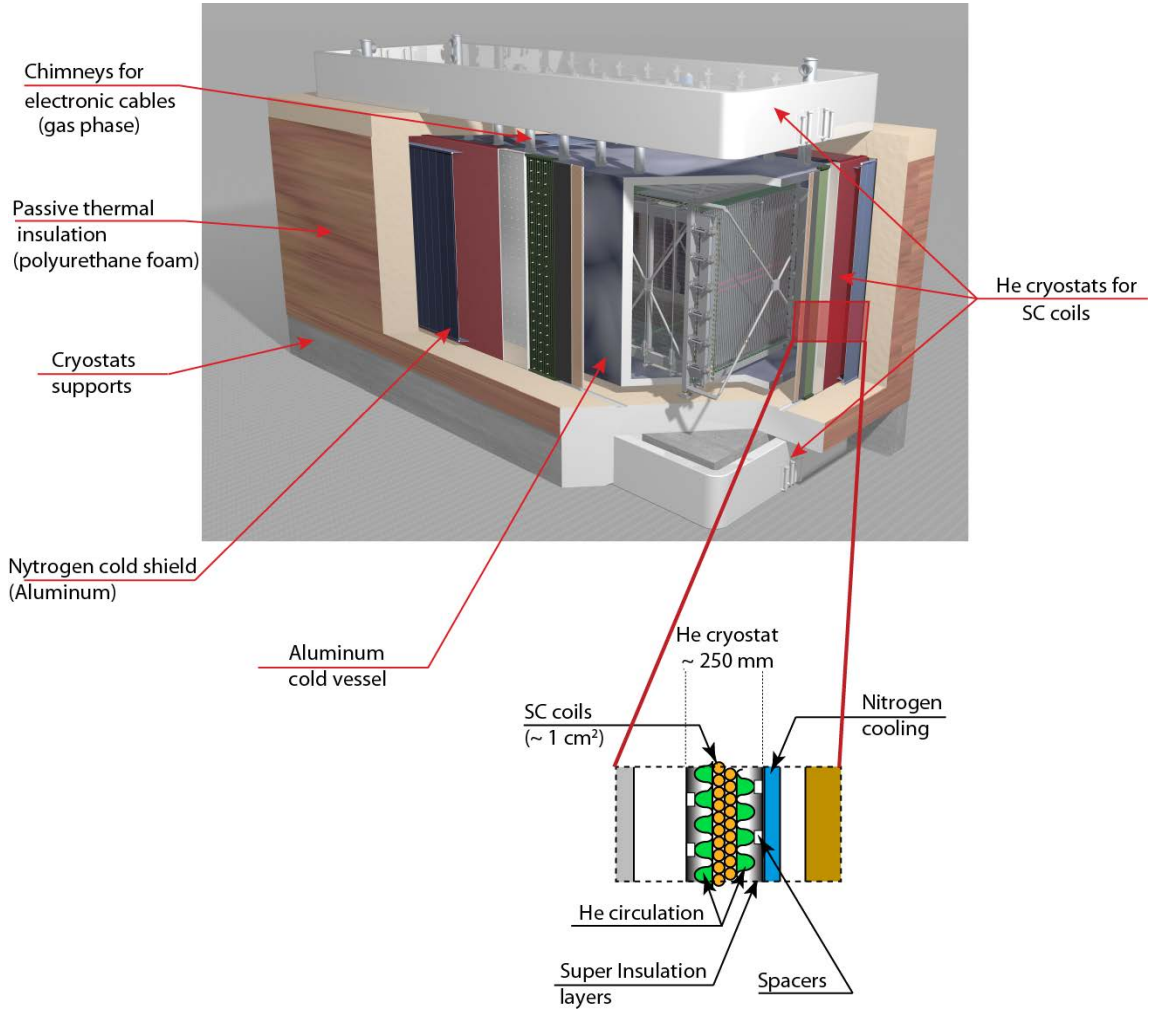

**Figure 17.** 3D General view of a possible implementation of a superconducting solenoid around the T150 detector. The main SC coils are located in the interstice between the detector Aluminum cold body and the Nitrogen cooling system. Two additional coils could be located above and below the passive insulation to make the magnetic field in LAr more uniform. A similar layout is foreseen for the T600.

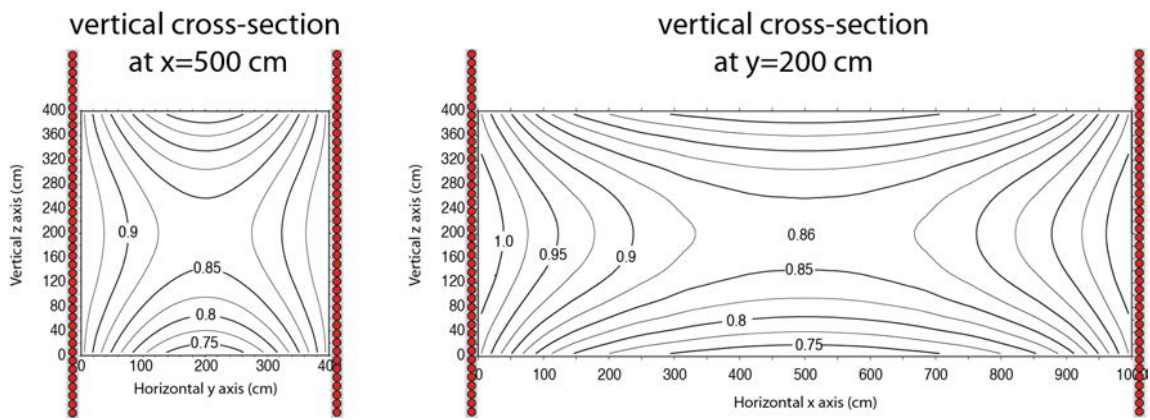

**Figure 18.** Equi-magnetic-field intensity lines along vertical cross-sections in the T150. The magnetic field direction is oriented along the vertical axis (z). The Superconducting coils (6 m tall) running horizontally around the cold aluminum vessel (see Figure 17 for details) are also sketched as red circles at the sides of the detectors. A very similar magnetic field map is also obtained for the T600.

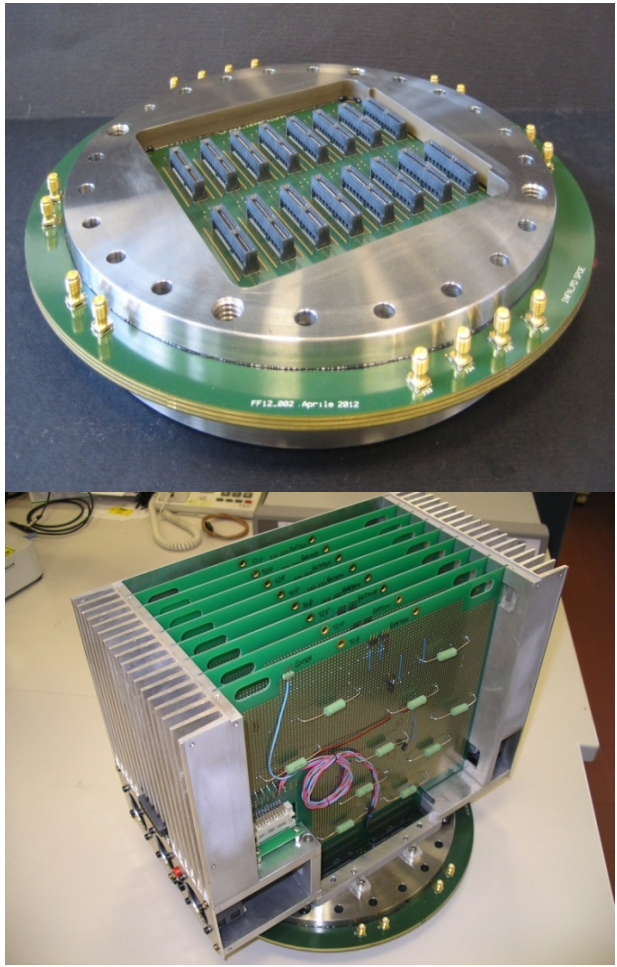

**Figure 19.** First prototypes of the flange and the electronics boards under development, with backplane integrated on the flange and power distribution on the auxiliary connectors on side bus.

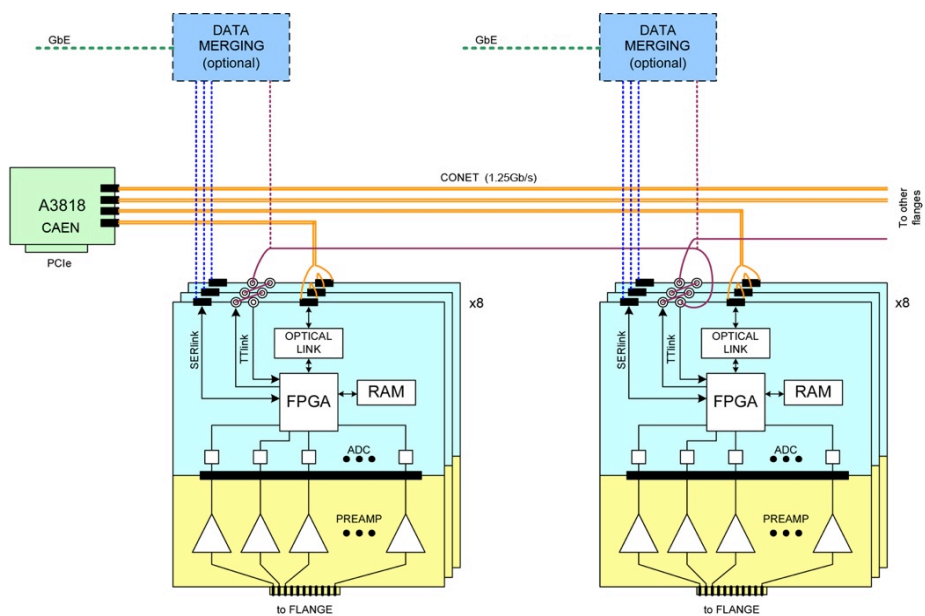

**Figure 20.** Block diagram of electronics housed on the flange.

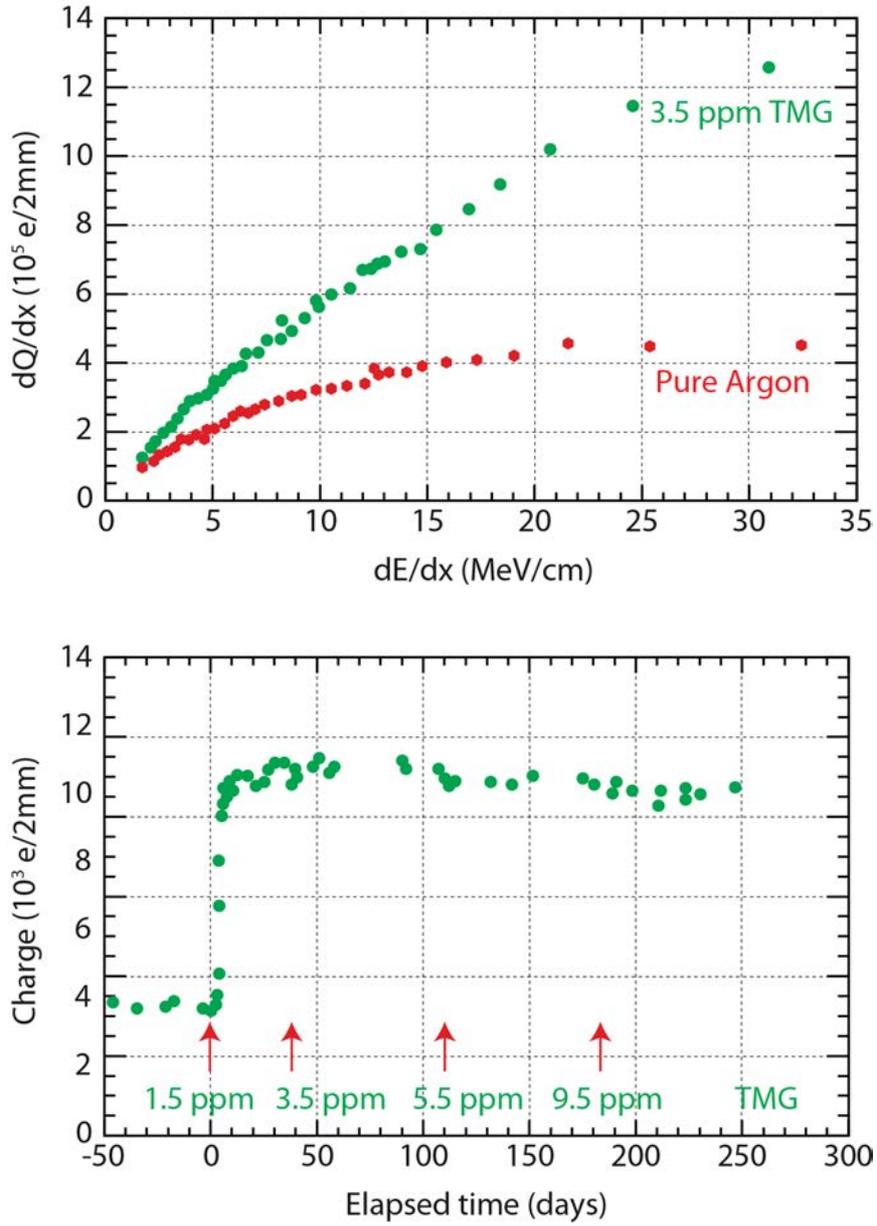

**Figure 21.** (Top) Collected charge versus deposited energy at an electric field of 200 V/cm, with pure Liquid Argon and with a TMG concentration of 3.5 ppm. (Bottom) Charge increase for mip tracks, showing full saturation already at TMG concentration of 1.3 ppm.

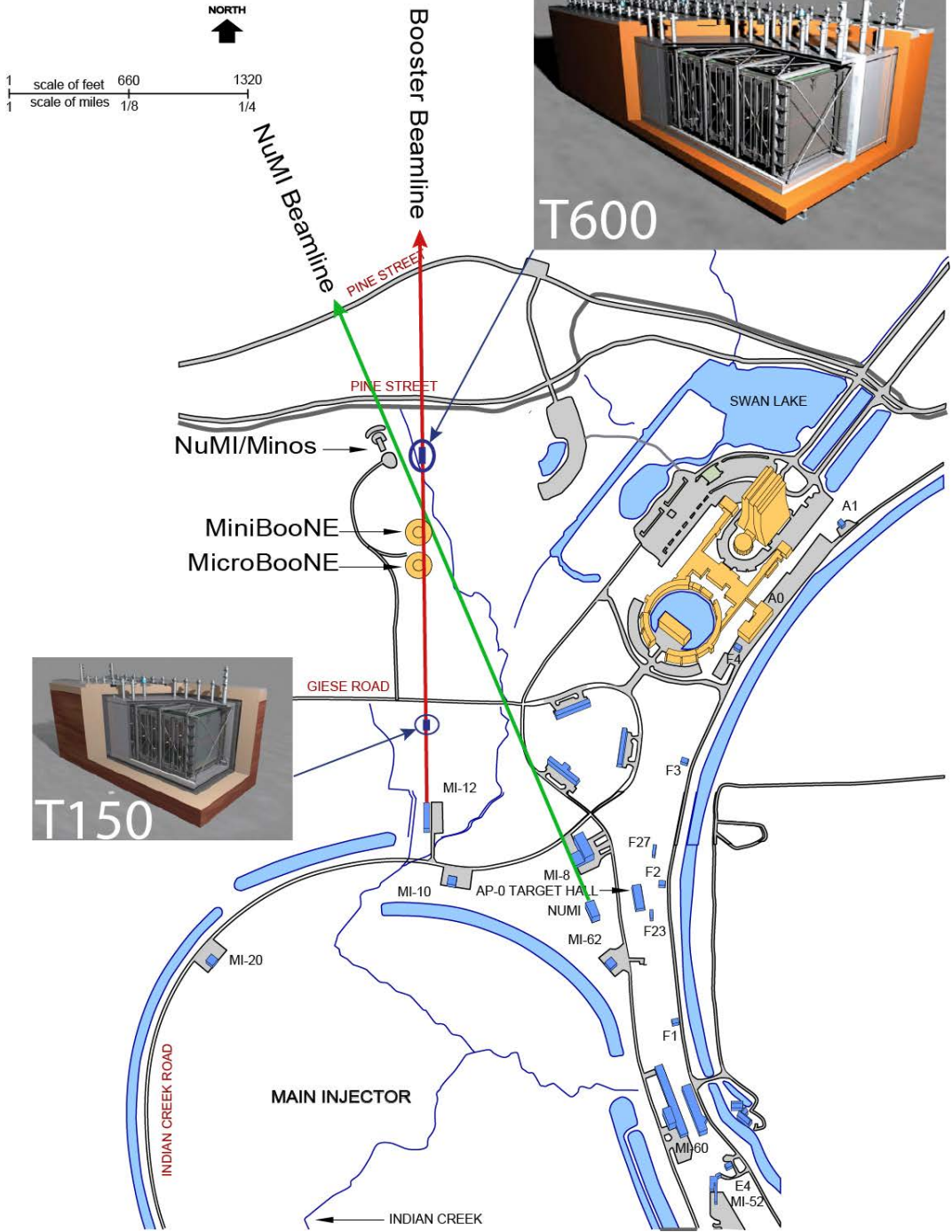

**Figure 22.** Proposed layout for the experiment at the FNAL Booster neutrino beam line.

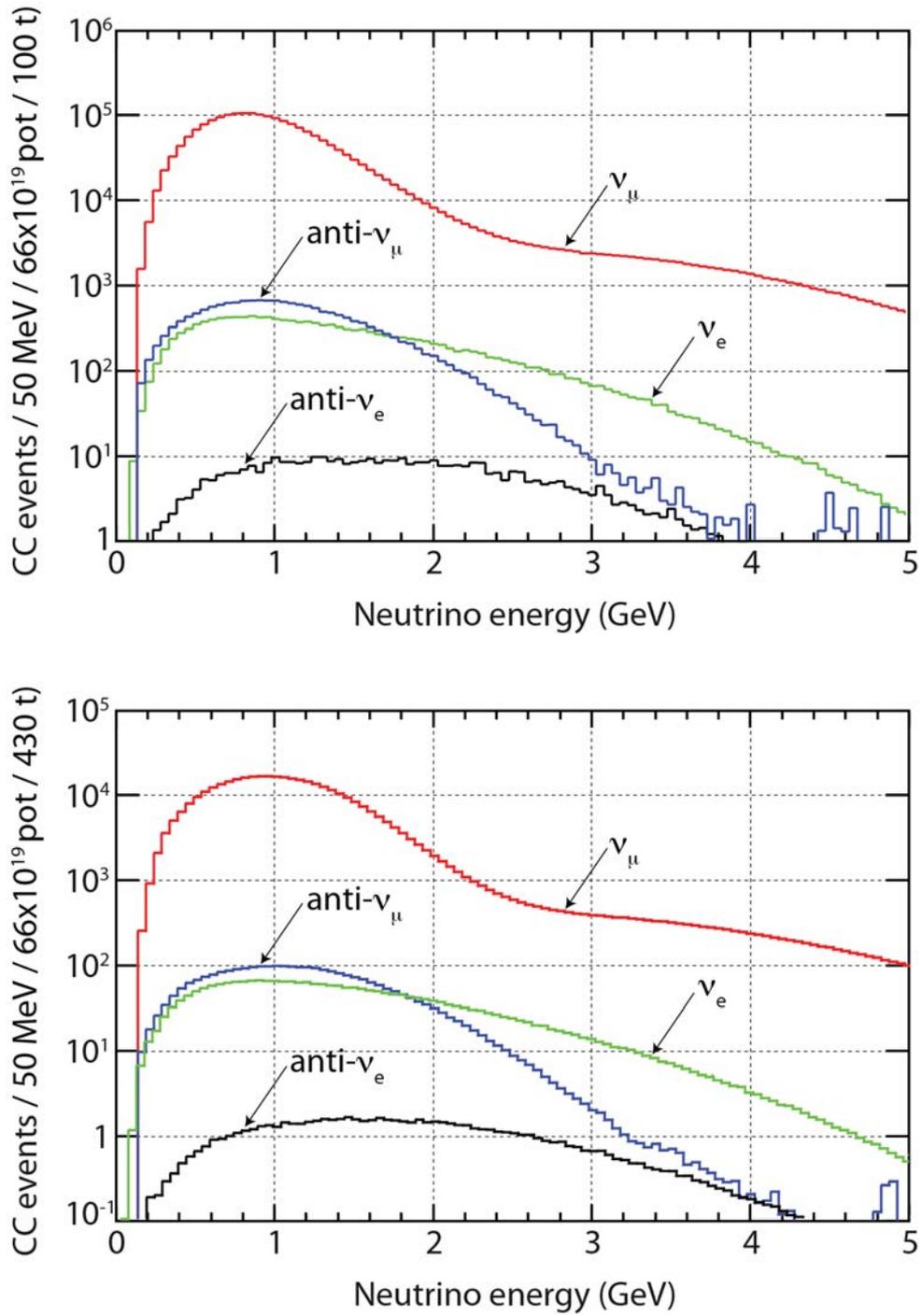

**Figure 23.** Neutrino CC event spectra and composition in case of positive focusing beam for $6.6 \times 10^{20}$ pot exposure. (Top) rates in the near detector (100 t fiducial mass at 150 m from target). (Bottom) Rates in the far detector (430 t fiducial mass at 700 m from target).

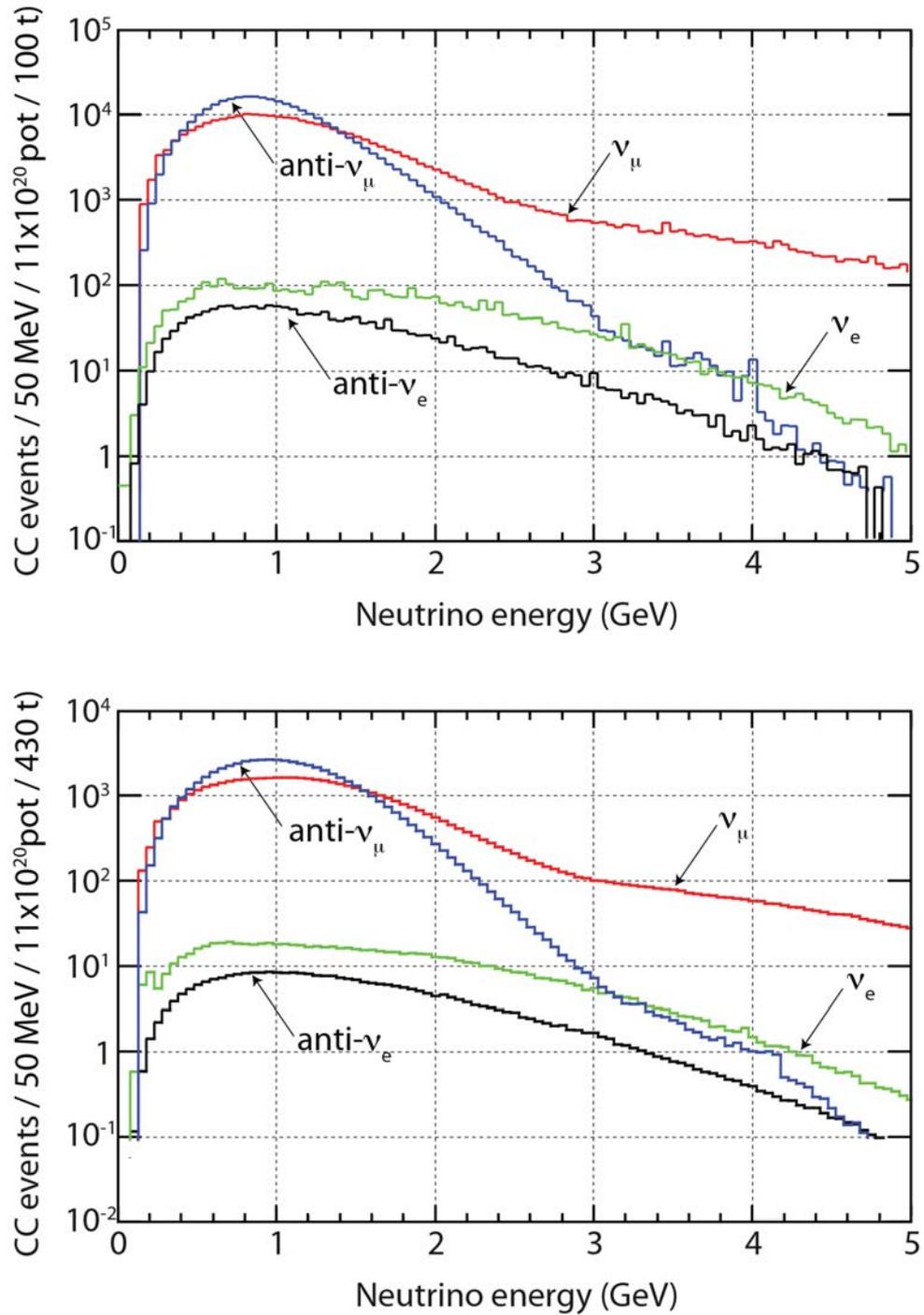

**Figure 24.** Neutrino CC event spectra and composition in case of negative focusing beam for $11 \times 10^{20}$ pot exposure. (Top) rates in the near detector (100 t fiducial mass at 150 m from target). (Bottom) Rates in the far detector (430 t fiducial mass at 700 m from target).

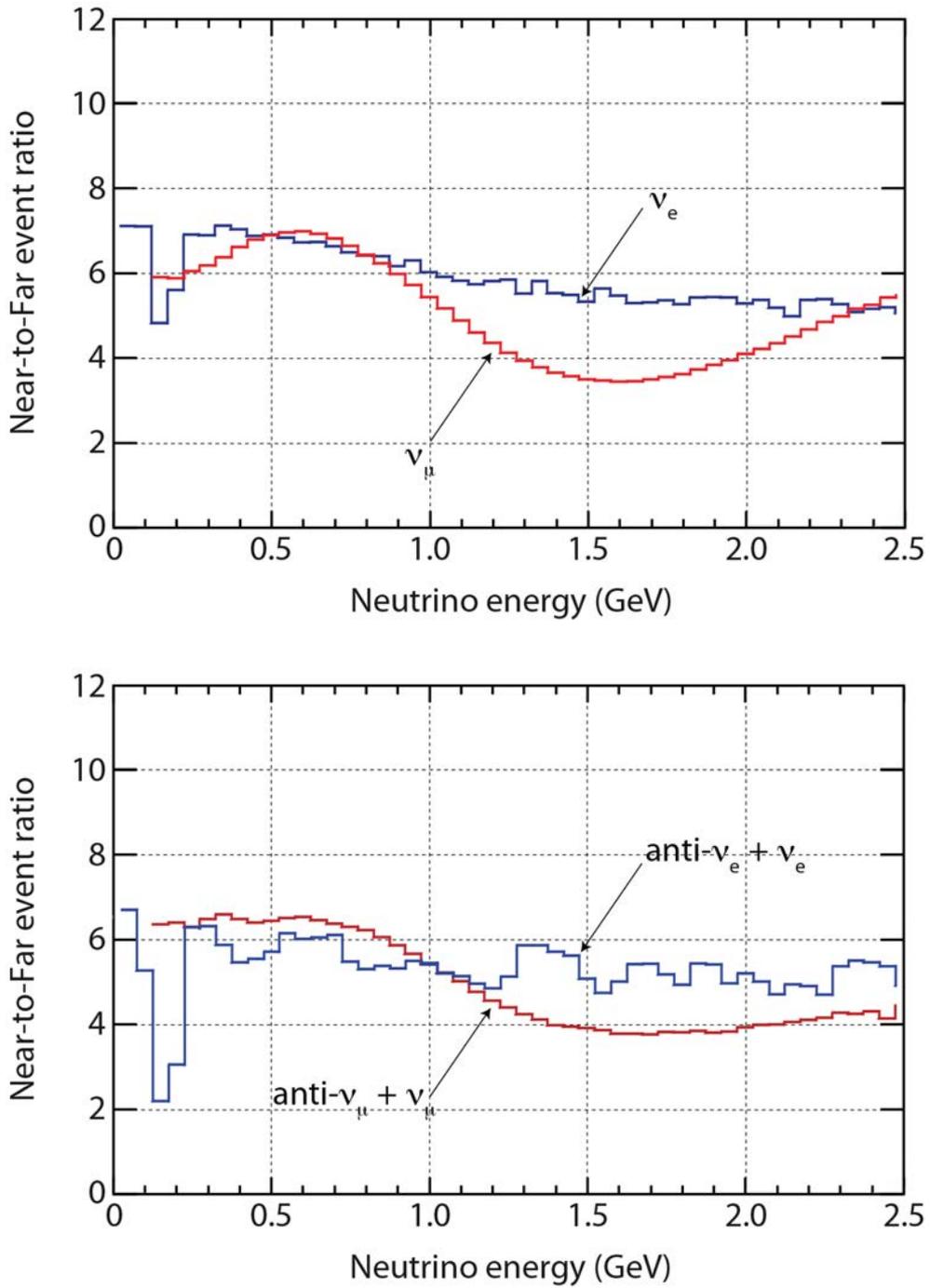

**Figure 25.** Near (150m) to Far (700m) detector CC event ratios for all neutrino components. (Top) Positive focusing. (Bottom) Negative focusing.

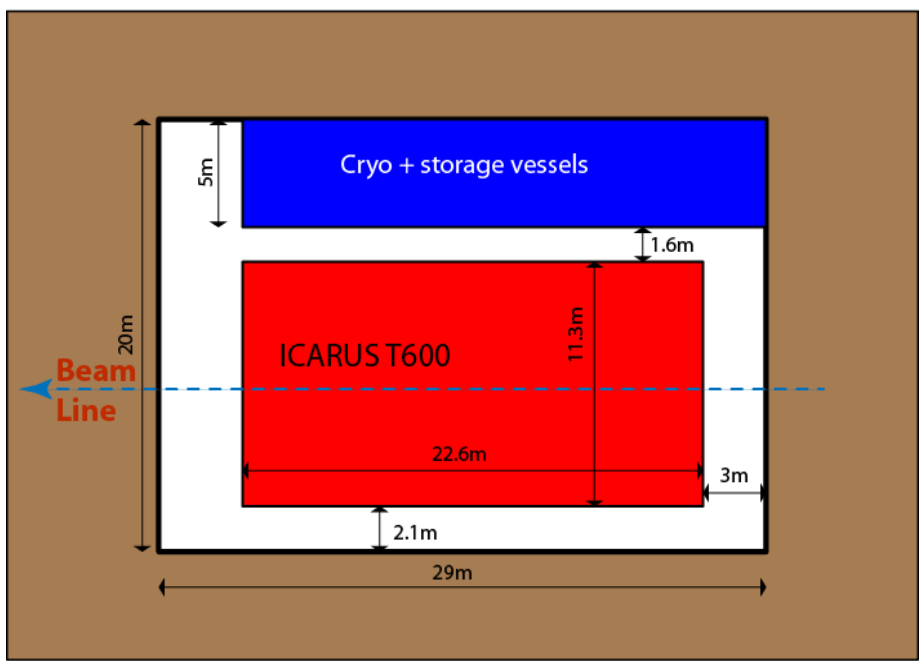

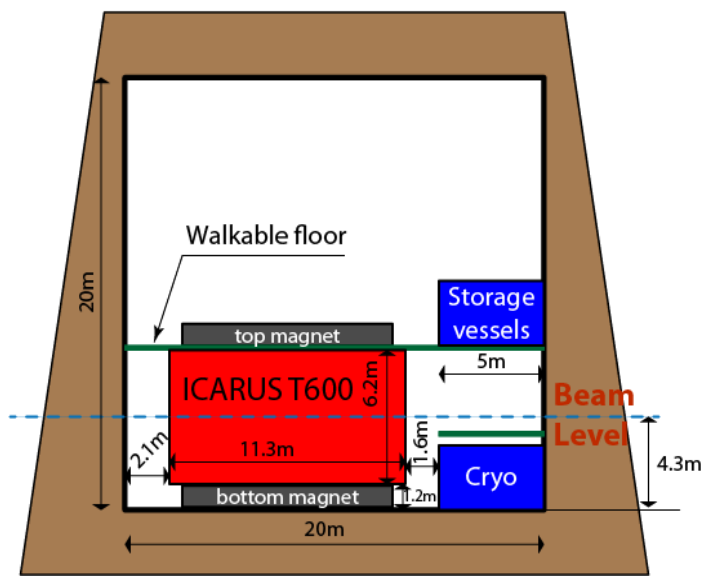

**Figure 26.** Schematics of the T600 installation at the Far location. The green line represents a walkable level capable to sustain the load of the electronic racks and of the LAr/LN2 storage vessels. An overburden of about 6 m of water equivalent is also required to filter the soft component of the cosmic rays.

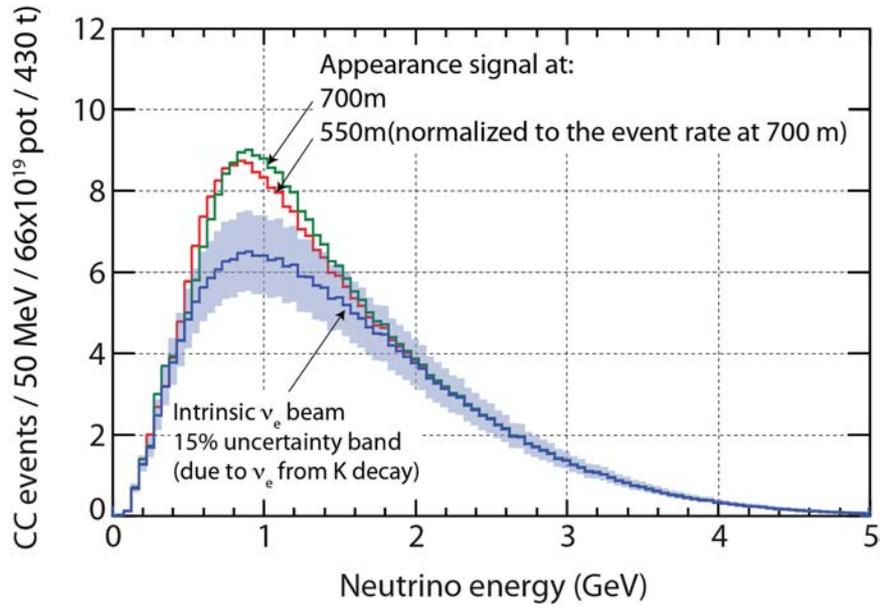

**Figure 27.** Examples of $\nu_e$ event spectra at the far detector in presence of $\nu_\mu \to \nu_e$ oscillations (positive polarity beam). The best fit parameters from the global analysis from Ref. [47] ($\Delta m^2 = 1.5 eV^2$ and $\sin^2(2\theta) = 0.0015$) have been used. The blue solid band represents the $1\sigma$ uncertainty on the $\nu_e$ intrinsic background at the far detector only, mainly due to systematic uncertainties on kaon production at the relevant energies estimated to be at the 30% level [26].

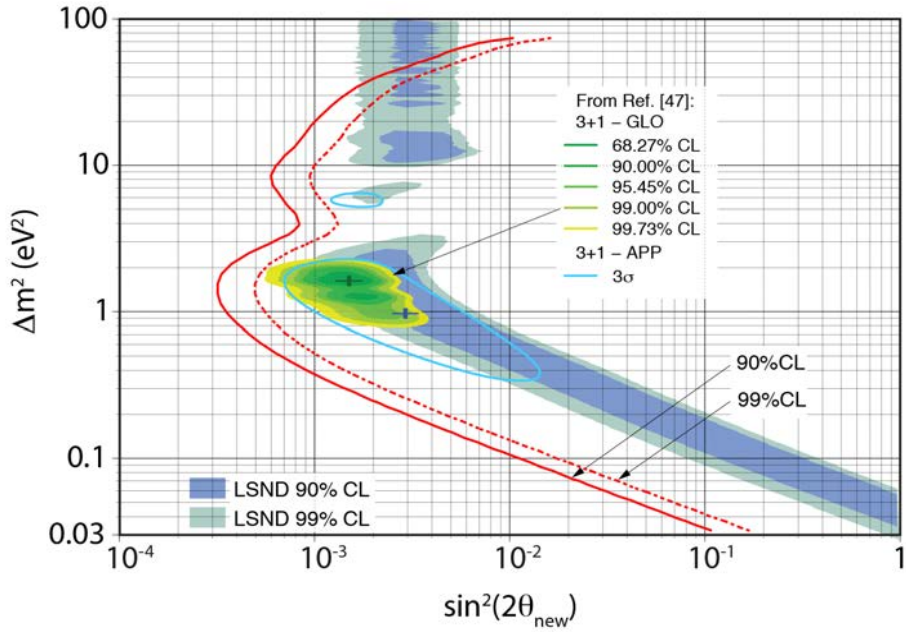

**Figure 28.** Expected sensitivity for $\nu_\mu \to \nu_e$ oscillations (positive polarity beam) for $6.6 \times 10^{20}$ pot compared with the LSND results. The predictions from the global 3+1 fit in [47] are also shown. A 4% uncorrelated systematic uncertainty has been assumed in the "Far" to "Near" $\nu_e$ ratio. The 90% C.L. corresponds to a $\Delta\chi^2 = 2.61$ (two sided 1 DOF $\chi^2$).

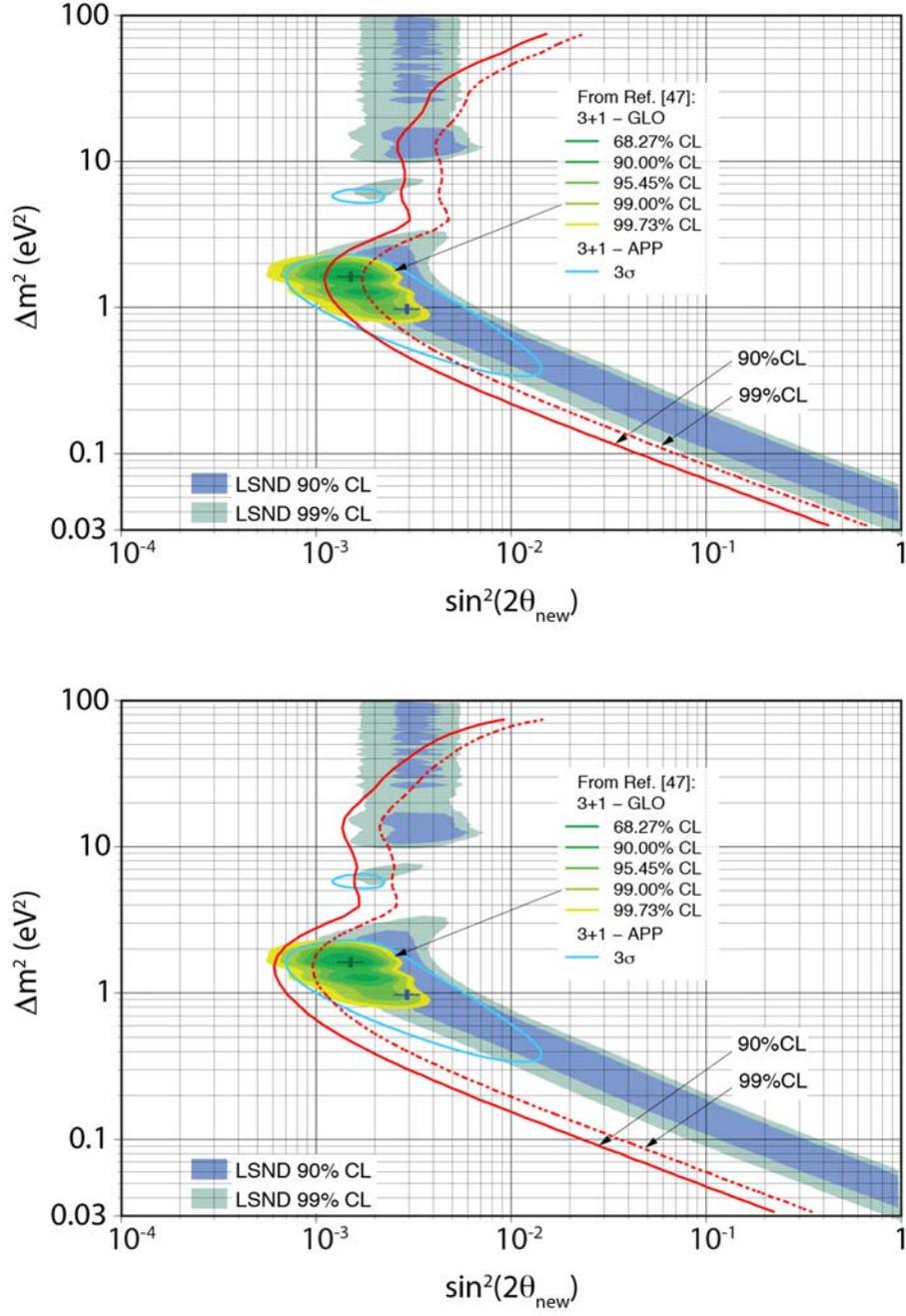

**Figure 29.** Expected sensitivity for anti-$\nu_\mu \to$ anti-$\nu_e$ oscillations (negative polarity beam) for $11 \times 10^{20}$ pot without (top) and with (bottom) the electron sign determination with magnetic field. The LSND results and the predictions from the global 3+1 fit in [47] are also shown. A 4% uncorrelated systematic uncertainty has been assumed in the "Far" to "Near" $\nu_e$ ratio. The 90% C.L. corresponds to a $\Delta\chi^2 = 2.61$ (two sided 1 DOF $\chi^2$).

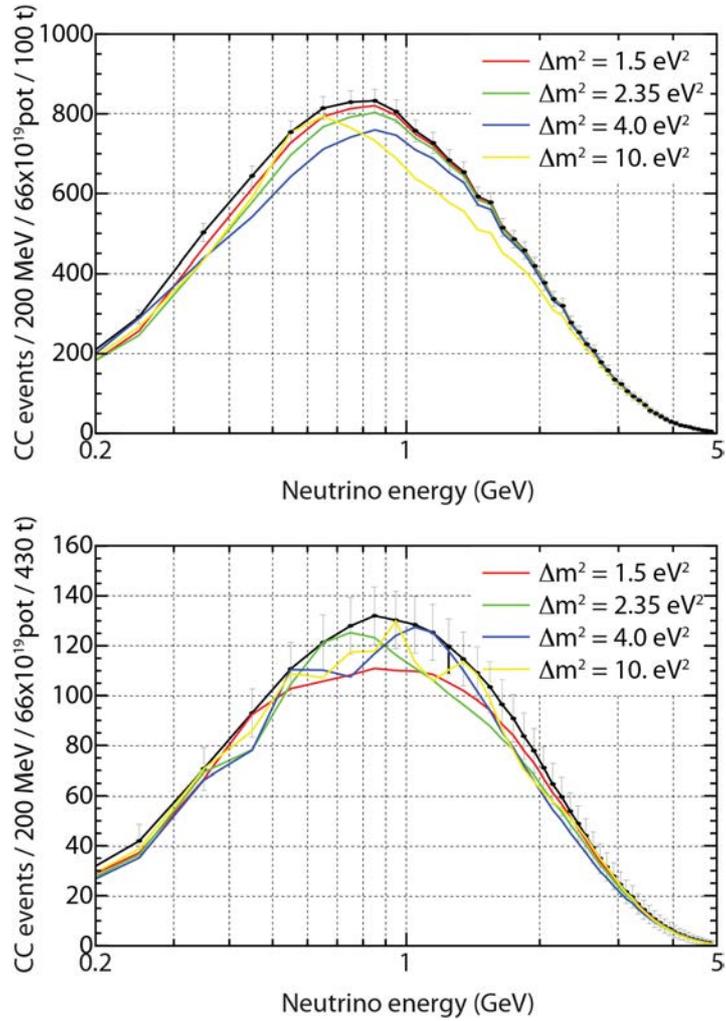

**Figure 30.** Energy distributions of the electron neutrino events in "Far" (top) and "Near" (bottom) detectors for a some $\Delta m^2$ values in the region with $\Delta m^2 > 1 eV^2$ and $\sin^2(2\theta) = 0.16$.

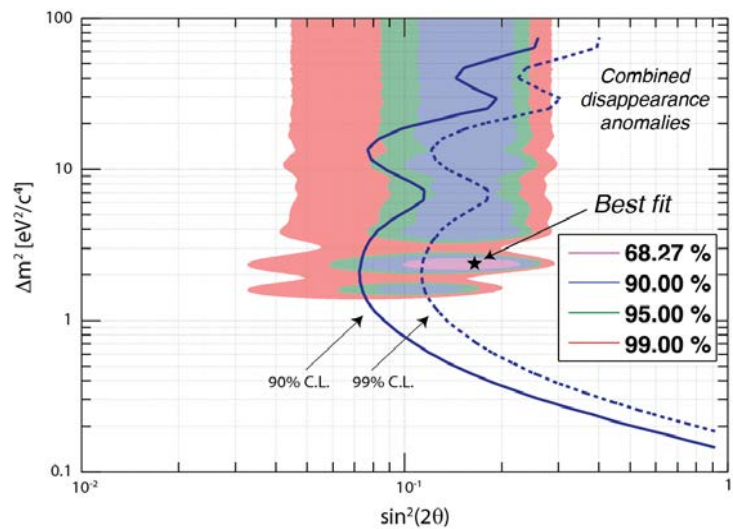

**Figure 31.** Sensitivity to $\nu_e$ disappearance in the 150m - 700m configuration and positive focusing for an exposure of $6.6\ 10^{20}$ pot at the Booster neutrino beam. As a comparison, the allowed region and best fit from Ref. [44] are also shown.

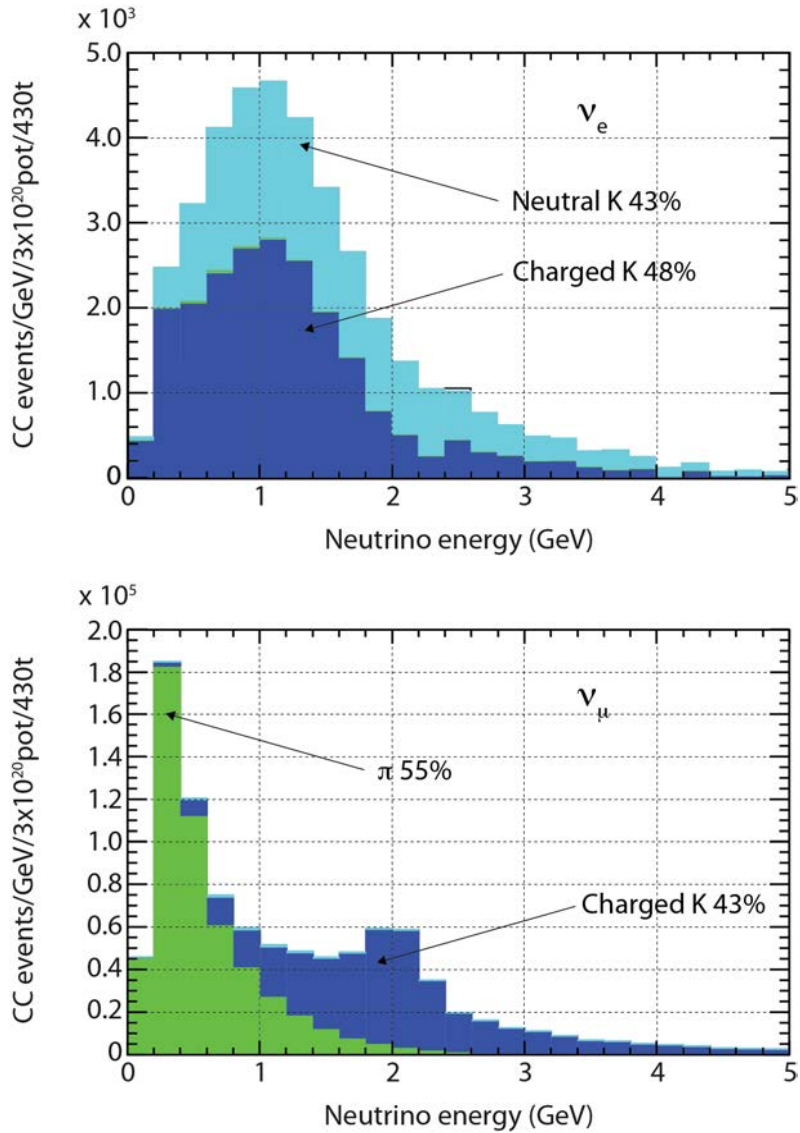

**Figure 32.** Neutrino rates in the T600 at the proposed "Far" location, ~105 mrad off-axis at ~880 m from the NuMI target for one year of exposure (~3 $10^{20}$ pot). (Top) Electron neutrino spectrum originates essentially from kaon decay and presents a broad peak covering the 0.5-2 GeV energy range. (Bottom) Muon neutrino components: a very low energy peak coming from pion decay is present as well as a peak at about 2 GeV due to kaon decay are clearly present.

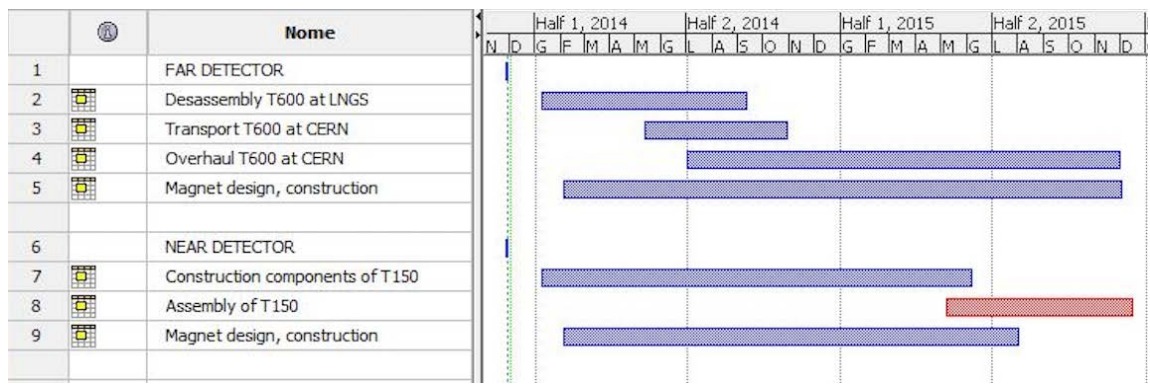

**Figure 33.** Time schedule for the overhauling of the T600 and the construction of the T150 detectors.

# 6  References.